\documentclass[aps,superscriptaddress,longbibliography,preprint]{revtex4-2}
\usepackage{graphicx}
\usepackage{amsmath,amssymb}

\usepackage{xcolor}

\newcommand{\fr}[2]{\frac{\displaystyle #1}{\displaystyle #2}}
\newcommand{\df}[2]{\frac{\displaystyle d#1}{\displaystyle d#2}}

\newcommand{\pf}[2]{\frac{\displaystyle \partial #1}{\displaystyle \partial #2}}

\newcommand{\Real}{\mathrm{Re}\:}
\newcommand{\Imag}{\mathrm{Im}\:}

\newcommand{\qed}{\hfill \rule{2.3mm}{2.3mm}}

\newtheorem{proposition}{Proposition}
\newtheorem{remark}{Remark}

\newcommand{\be}{\begin{equation}}
\newcommand{\ee}{\end{equation}}

\begin{document}

\title{Activity patterns in ring networks of quadratic integrate-and-fire neurons with synaptic
and gap junction coupling}

\author{Oleh E. Omel'chenko}
\email{omelchenko@uni-potsdam.de}
\address{University of Potsdam,
Institute of Physics and Astronomy,
Karl-Liebknecht-Str. 24/25,
14476 Potsdam,
Germany}
%ORCID: 0000-0003-0526-1878

\author{Carlo R. Laing}
\email{c.r.laing@massey.ac.nz}
\address{School of Mathematical and Computational Sciences, 
Massey University, Private Bag 102-904 NSMC, Auckland, New Zealand}
%ORCID: 0000-0002-6086-2978

\date{\today}
%\pacs{05.45.Xt, 05.45.Ac}

\begin{abstract}
We consider a ring network of quadratic integrate-and-fire neurons with nonlocal synaptic
and gap junction coupling. The corresponding neural field model supports solutions such as 
standing and travelling waves, and also lurching waves. We show that many of these solutions
satisfy self-consistency equations which can be used to follow them as parameters are varied.
We perform numerical bifurcation analysis of the neural field model, concentrating on
the effects of varying gap junction coupling strength. Our methods are generally applicable
to a wide variety of networks of quadratic integrate-and-fire neurons.
\end{abstract}

\maketitle

\section{Introduction}
The collective behaviour of spatially extended networks of neurons is a topic of ongoing
interest~\cite{ermter10,coowed23,coogra14,bre12,ByrRNC2022,erm98}. While it is possible
to simulate large networks of model neurons~\cite{wimnyk14,horoph97}, continuum level
descriptions (neural field models) often provide more potential for mathematical analysis.
Early neural field models were phenomenological~\cite{ama77,laitro03,wilcow73,coo05} but more
recently models derived rigorously from infinite networks of spiking neurons have become
available~\cite{lai15,omelai22,lai14a,LaiO2023,esnrox17,schavi20,byravi19}.

One type of solution of a neural field model is a ``bump'' --- a spatially localised group
of active neurons. Such solutions are thought to be relevant to working memory~\cite{Combru00} and
the head direction system~\cite{xiehah02,khofie22}. 
Also important are travelling waves~\cite{pinerm01a,breweb12,huatro04,osarub02}.
These are thought to be relevant for modelling epilepsy and migraines,
see for example~\cite{brecar15}.
We are often interested in when and how such solutions lose stability or are destroyed in bifurcations,
and what patterns are stable beyond such bifurcations. For example, a bump may start to 
``breathe''~\cite{folbre04} or a travelling wave may no longer have a fixed profile.

In this paper we study a ring network of quadratic integrate-and-fire (QIF) neurons coupled
both synaptically and via gap junctions. We analyse the continuum network,
whose dynamics are described by a neural field model. In similar previous work we considered
networks of theta neurons with just synaptic coupling, concentrating
on only time-periodic solutions~\cite{LaiO2023}. (The QIF neuron
with infinite threshold and reset is equivalent to a theta neuron~\cite{monpaz15}.)
In earlier work we also studied the stationary states
of a ring network of synaptically coupled theta neurons~\cite{omelai22}.
On the other hand, models similar (but not identical) to ours
have been the focus of other reasearchers.
Byrne et al. studied networks of QIF neurons with both
synaptic and gap junction coupling, but their synaptic coupling incorporated propagation
delays and the gap junction coupling was purely local~\cite{ByrRNC2022}. Their neural field
model was only valid in the long wavelength limit. Schmidt and Avitabile considered a ring
network of QIF neurons with nonlocal synaptic coupling and analysed both steady states and
time dependent solutions which arose as the result of periodic forcing~\cite{schavi20}.
Esnaola-Acebes et al. studied a similar model, focussing on the decaying oscillatory modes
that such models show~\cite{esnrox17}. Byrne et al. also considered a ring network of QIF neurons
with synaptic coupling, performing a largely numerical study of some of the possible types of
solutions~\cite{byravi19}.

Our model is probably most similar to that of~\cite{ByrRNC2022}, but one of our contributions 
is to show
that some solutions of interest can be analysed using a self-consistency approach that has
previously been used only for networks of theta neurons or Kuramoto and Winfree
oscillators~\cite{LaiO2023,ome23,batcle23}. The structure of the paper is as follows.
In Sec.~\ref{sec:model} we present the model and describe the types of patterns typically
seen when numerically solving it. In Sec.~\ref{sec:phen} we choose two values of the
synaptic coupling strength and vary the strength of the gap junction coupling, observing
 transitions between various types of solutions. In Sec.~\ref{sec:bif} we present analytical
methods which can be used to describe most of the solutions observed in Sec.~\ref{sec:phen}.
Section~\ref{sec:results} shows the results of implementing the methods described in
Sec.~\ref{sec:bif} and includes a discussion of the relative efficiencies of several
alternative methods. We conclude in Sec.~\ref{sec:disc}, and the Appendix contains a number
of useful results about the solutions of the complex Riccati equation.

\section{Model and types of solutions}
\label{sec:model}

Let us consider a spatially extended network
of $N$ quadratic integrate-and-fire (QIF) neurons
with both gap-junction and pulsatile synaptic coupling:
\begin{equation}
\df{V_j}{t} = \eta_j + V_j^2 + \kappa_\mathrm{v} \ell \sum\limits_{k=1}^N W_\mathrm{v}\left( |j-k| \ell \right) ( V_k - V_j ) 
+ \kappa_\mathrm{s} \ell \sum\limits_{k=1}^N \sum\limits_{m\in\mathbb{Z}} W_\mathrm{s}\left( |j-k| \ell \right) \delta\left(t - T_k^m \right).
\label{Eq:Nonlocal}
\end{equation}
Here, the spiking events of the $j$th neuron $T_j^m$ are determined
by the reset condition: if $V_j\to+\infty$ for $t\nearrow T_j^m$,
then $V_j\to-\infty$ for $t\searrow T_j^m$.

We assume that the network is organized as a one-dimensional array
with periodic boundary conditions and distance-dependent connectivity
defined by coupling functions $W_\mathrm{v}$ and $W_\mathrm{s}$.
The neurons differ from each other only in the excitability parameters~$\eta_j$,
which are chosen randomly and independently from a Lorentzian distribution
$$
g(\eta) = \fr{1}{\pi} \fr{\gamma}{(\eta - \eta_0)^2 + \gamma^2}\quad\mbox{with}\quad \eta_0\in\mathbb{R}\quad\mbox{and}\quad\gamma > 0.
$$
The coupling functions $W_\mathrm{v}$ and $W_\mathrm{s}$
are assumed to be even, continuous and periodic
$$
W_\mathrm{v}\left( (k+N) \ell \right) = W_\mathrm{v}\left( k \ell \right),\qquad
W_\mathrm{s}\left( (k+N) \ell \right) = W_\mathrm{s}\left( k \ell \right),
$$
and the intensity of interactions between neurons is controlled
by two scalar coupling strengths $\kappa_\mathrm{v}$ and $\kappa_\mathrm{s}$.
Moreover, for the sake of simplicity of the analytical consideration below,
it is convenient to assume $\ell = 2\pi / N$.
In this case, functions $W_\mathrm{v}(x)$ and $W_\mathrm{s}(x)$ are $2\pi$-periodic
and can be represented in the form of Fourier series
$$
W_\mathrm{v}(x) = \sum\limits_{m=-\infty}^\infty \hat{W}_{\mathrm{v},m} e^{i m x}
\quad\mbox{and}\quad
W_\mathrm{s}(x) = \sum\limits_{m=-\infty}^\infty \hat{W}_{\mathrm{s},m} e^{i m x}
$$
with coefficients
\begin{equation}
\hat{W}_{\mathrm{v},m} = \fr{1}{2\pi} \int_{-\pi}^{\pi} W_\mathrm{v}(x) e^{-i m x} dx
\quad\mbox{and}\quad
\hat{W}_{\mathrm{s},m} = \fr{1}{2\pi} \int_{-\pi}^{\pi} W_\mathrm{s}(x) e^{-i m x} dx.
\label{Fourier:W}
\end{equation}
Note that for even functions $W_\mathrm{v}(x)$ and $W_\mathrm{s}(x)$,
the coefficients $\hat{W}_{\mathrm{v},m}$ and $\hat{W}_{\mathrm{s},m}$ are real and therefore
$$
\hat{W}_{\mathrm{v},m} = \fr{1}{2\pi} \int_{-\pi}^{\pi} W_\mathrm{v}(x) \cos(m x) dx
\quad\mbox{and}\quad
\hat{W}_{\mathrm{s},m} = \fr{1}{2\pi} \int_{-\pi}^{\pi} W_\mathrm{s}(x) \cos(m x) dx.
$$

It is well-known~\cite{lai15,ByrRNC2022,piedev19,monpaz20,monpaz15} that in the continuum limit $N\to\infty$,
the long-term dynamics of system~(\ref{Eq:Nonlocal})
can be described by a neural field equation
\begin{equation}
\pf{u}{t} = \gamma - \kappa_\mathrm{v} u + i \left[ \eta_0 + \kappa_\mathrm{v} \mathcal{K}_\mathrm{v} \Imag(u) + \fr{\kappa_\mathrm{s}}{\pi} \mathcal{K}_\mathrm{s} \Real(u) - u^2 \right]
\label{Eq:MeanField}
\end{equation}
for a complex-valued function $u = u(x,t)$,
where $\mathcal{K}_\mathrm{v}$ and $\mathcal{K}_\mathrm{s}$
denote two integral operators
$$
( \mathcal{K}_\mathrm{v} \varphi )(x) = \int_0^{2\pi} W_\mathrm{v}(x - y) \varphi(y) dy
\quad\mbox{and}\quad
( \mathcal{K}_\mathrm{s} \varphi )(x) = \int_0^{2\pi} W_\mathrm{s}(x - y) \varphi(y) dy.
$$
Note that the real and imaginary parts of the solution $u(x,t)$ can be related to the local firing rate
\begin{equation}
R(x,t) = \fr{1}{\pi} \Real u(x,t)
\label{Def:R}
\end{equation}
and the local mean field potential
\begin{equation}
V(x,t) = \Imag u(x,t).
\label{Def:V}
\end{equation}
Due to this interpretation, only those solutions
that satisfy $\Real(u)\ge 0$ are physically meaningful.
Moreover, it can be explicitly shown that the property $\Real(u)\ge 0$
is preserved by the dynamics of Eq.~(\ref{Eq:MeanField}).
Therefore, we focus on this invariant set below.

Note also that
the mapping
\begin{equation}
z(x,t) = \fr{1 - \overline{u}}{1 + \overline{u}},
\label{Def:z}
\end{equation}
where overline denotes the complex conjugate,
transforms the function $u$ into an analog of the Kuramoto local order parameter, $z$; the
dynamics can be written equally-well in terms of $z$~\cite{lai15,ByrRNC2022}.

In the examples below, we choose $W_\mathrm{v}(x)$ in the form of a Gaussian function
\begin{equation}
W_\mathrm{v}(x) = \fr{1}{\sqrt{2\pi}\sigma} e^{-x^2/(2\sigma^2)}
\quad\mbox{with}\quad \sigma = 0.1,
\label{Wv:Example}
\end{equation}
and $W_\mathrm{s}(x)$ in the form of a Mexican-hat function which is positive for small $|x|$
and negative for larger $|x|$:
\begin{equation}
W_\mathrm{s}(x) = \fr{1}{\sqrt{2\pi}\sigma_1} e^{-x^2/(2\sigma_1^2)}
- \fr{1}{\sqrt{2\pi}\sigma_2} e^{-x^2/(2\sigma_2^2)}
\quad\mbox{with}\quad \sigma_1 = 0.5,\: \sigma_2 = 1.
\label{Ws:Example}
\end{equation}
More precisely, the above functions determine the coupling only for $|x| \le \pi$,
while we use their $2\pi$-periodic extensions for other values of $x$. Note that the
width of the gap junction coupling kernel, $W_\mathrm{v}$, is much smaller than 
that of the synaptic coupling kernel, $W_\mathrm{s}$, since gap junctional coupling
is typically more localised than synaptic coupling~\cite{lai15}. 

In numerical simulations for the spatially discretized version of Eq.~(\ref{Eq:MeanField}),
we usually encountered the following five types of stable states:

\begin{enumerate}

\item  Uniform states, i.e. constant solutions of Eq.~(\ref{Eq:MeanField})
that do not depend on either $x$ or $t$.

\item Stationary states, i.e. time-independent solutions of the form $u = a(x)$ for
some $2\pi$-periodic function $a$.

\item Traveling waves, i.e. solutions of the form $u = a(x - s t)$,
which are stationary in a frame moving with a constant speed $s\ne 0$.

\item Standing waves, i.e. solutions $u = a(x,t)$
satisfying the periodicity condition $a(x,t+T) = a(x,t)$ with some $T > 0$.

\item Lurching waves, i.e. solutions $u = a(x,t)$
satisfying the shifted periodicity condition $a(x+\chi,t+T) = a(x,t)$
with some $\chi\ne 0$ and $T > 0$. Such waves have been found previously in neural
models~\cite{wascis10,coo03,golerm99,golerm00,golerm01,yewter01}.
\end{enumerate}

In this paper we analyse all of these types of solutions using a mixture of numerical
and analytical methods. In Sec.~\ref{sec:phen} we show examples of these types of
solution and characterise some aspects of them as the parameter $\kappa_\mathrm{v}$
(the strength of gap junction coupling) is varied. We choose to vary this parameter
as the influence of gap junction coupling in neural field models has only recently
been considered~\cite{lai15,monpaz20,piedev19}.
In Sec.~\ref{sec:bif} we undertake
bifurcation analysis of the various types of solution, sometimes describing them using a recently
presented method that can be used to efficiently characterise solutions using self-consistency
arguments~\cite{ome23,ome22,LaiO2023,batcle23}.

\section{Phenomenology}
\label{sec:phen}

To determine the typical attractors of the neural field equation~(\ref{Eq:MeanField}),
we discretized it on a uniform spatial grid of $1024$ nodes
and performed numerical simulations for various parameters
$\kappa_\mathrm{s}$ and $\kappa_\mathrm{v}$.
Other parameters were chosen as $\eta_0 = 1$ and $\gamma = 0.5$,
by analogy with a similar model considered by Byrne et al. in~\cite{ByrRNC2022}.
Our observations are summarized below, where we provide a comprehensive overview
of two representative cases $\kappa_\mathrm{s} = 10$ and $\kappa_\mathrm{s} = 20$.

\subsection{Case $\kappa_\mathrm{s} = 10$}
\label{sec:phen10}

\begin{figure}[h]
\includegraphics[width=0.45\textwidth]{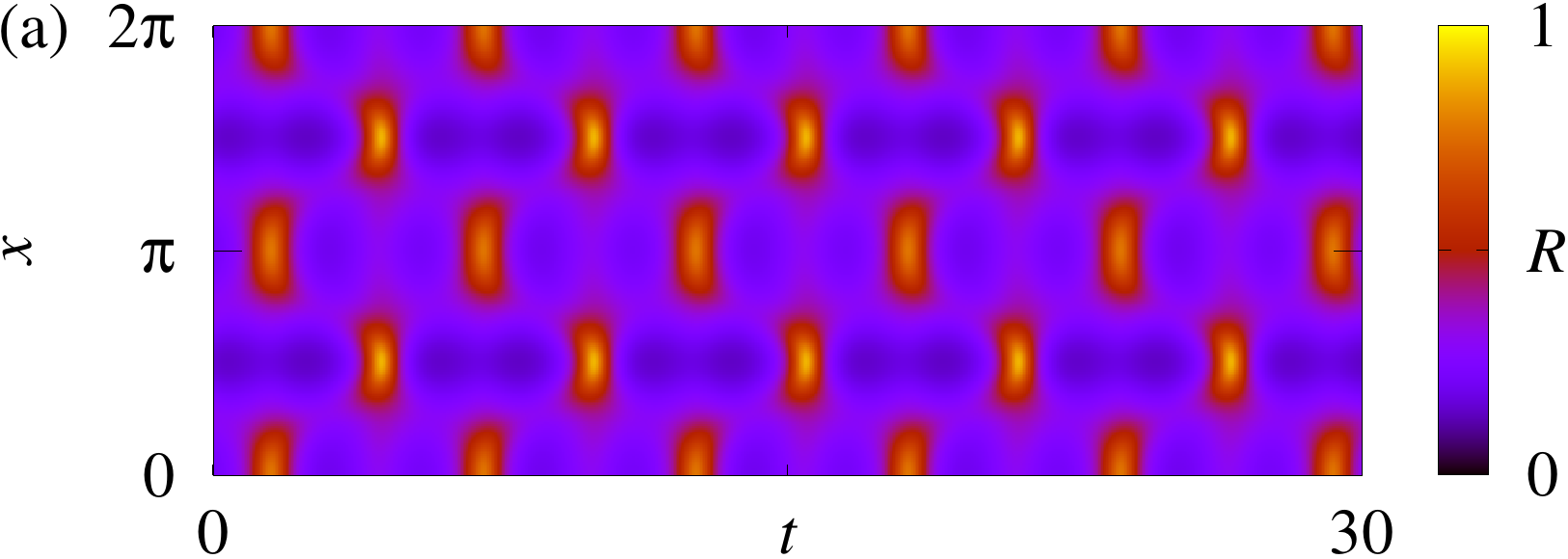}\hspace{5mm}
\includegraphics[width=0.45\textwidth]{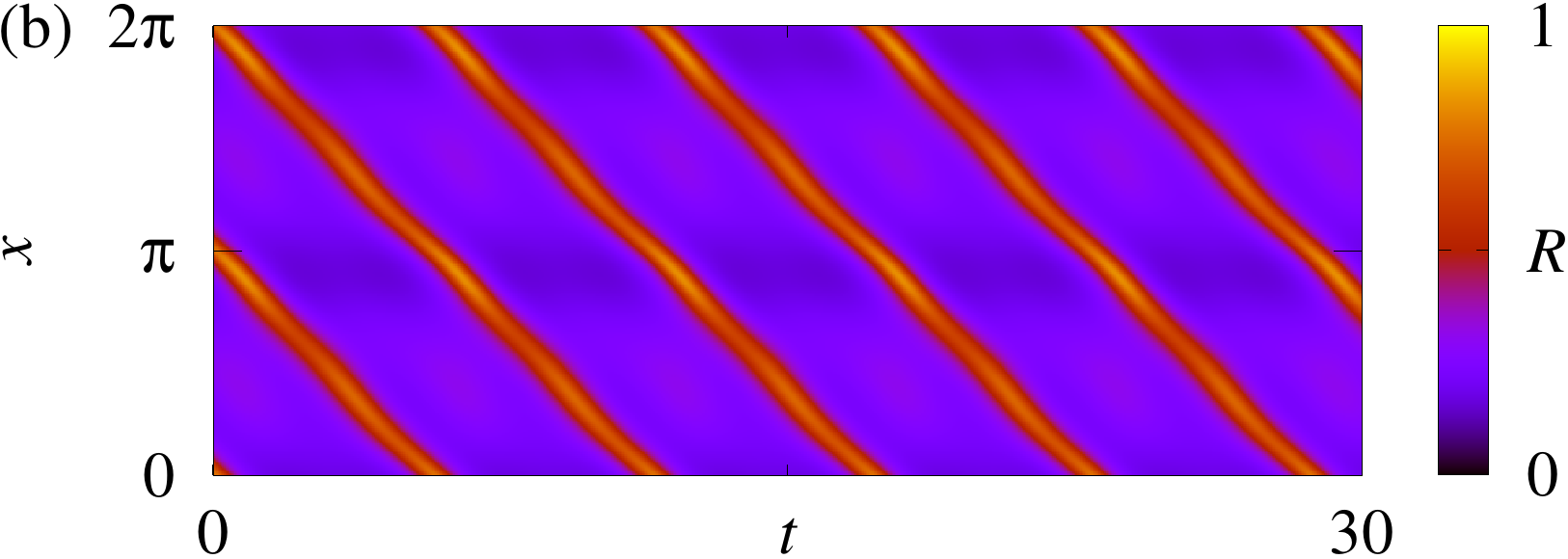}
\caption{(a) A standing wave solution of Eq.~(\ref{Eq:MeanField})
and (b) a coexisting travelling wave solution.
For both states, the local firing rate $R$,
calculated by formula~(\ref{Def:R}), is shown.
Parameters: $\kappa_\mathrm{s} = 10$, $\kappa_\mathrm{v} = 1$,
$\eta_0 = 1$ and $\gamma = 0.5$.}
\label{Fig:SW:TW}
\end{figure}

Choosing $\eta_0 = 1$, $\gamma = 0.5$,
$\kappa_\mathrm{s} = 10$ and $\kappa_\mathrm{v} = 1$
and running the numerical simulations for Eq.~(\ref{Eq:MeanField})
with various initial conditions, we found two coexisting spatio-temporal patterns;
see Fig.~\ref{Fig:SW:TW}.
One of them had the appearance of a periodically oscillating standing wave,
while the other had the appearance of a travelling wave, although the wave does not travel
with a constant profile. Note that standing and travelling waves have been observed in several
other neural field models~\cite{LaiO2023,byravi19,ByrRNC2022}.

%%%%%%%%%%%%%%%%%%%%%%%%%%%%%%%%%%%%%%%%%%%%%%%%
%%%%%%%%%%%%%%%%%%%%%%%%%%%%%%%%%%%%%%%%%%%%%%%%

\begin{figure}[h]
\includegraphics[width=0.45\textwidth]{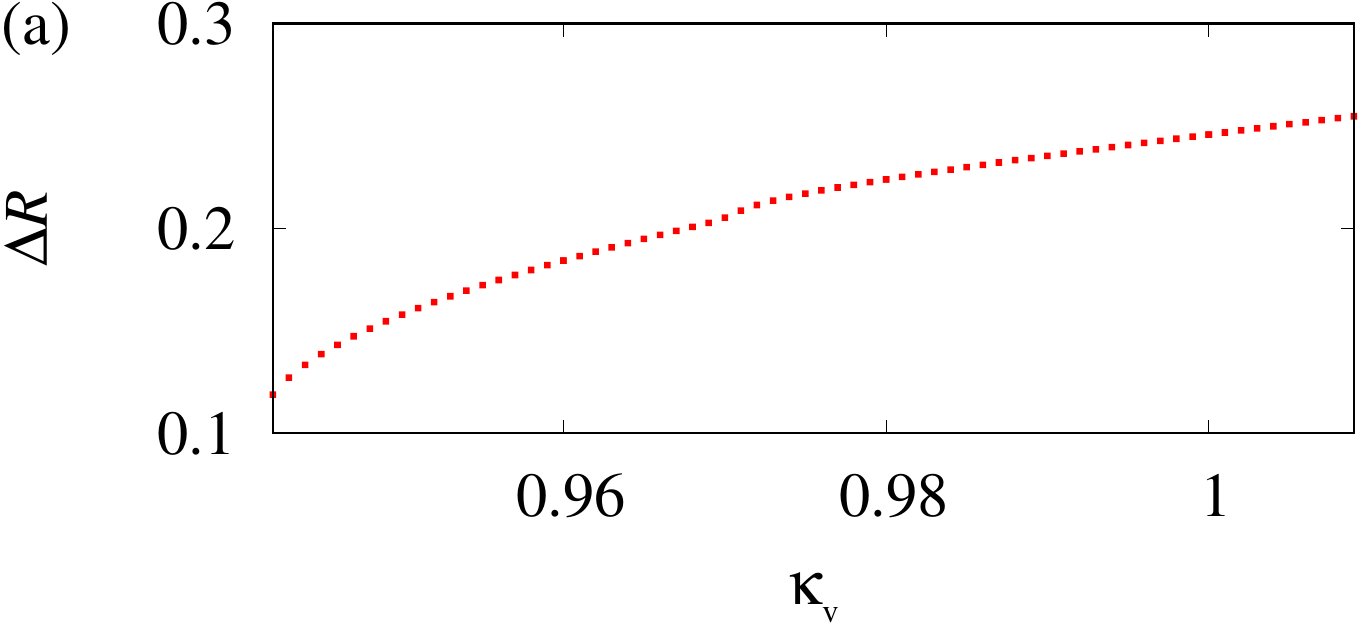}\hspace{8mm}
\includegraphics[width=0.45\textwidth]{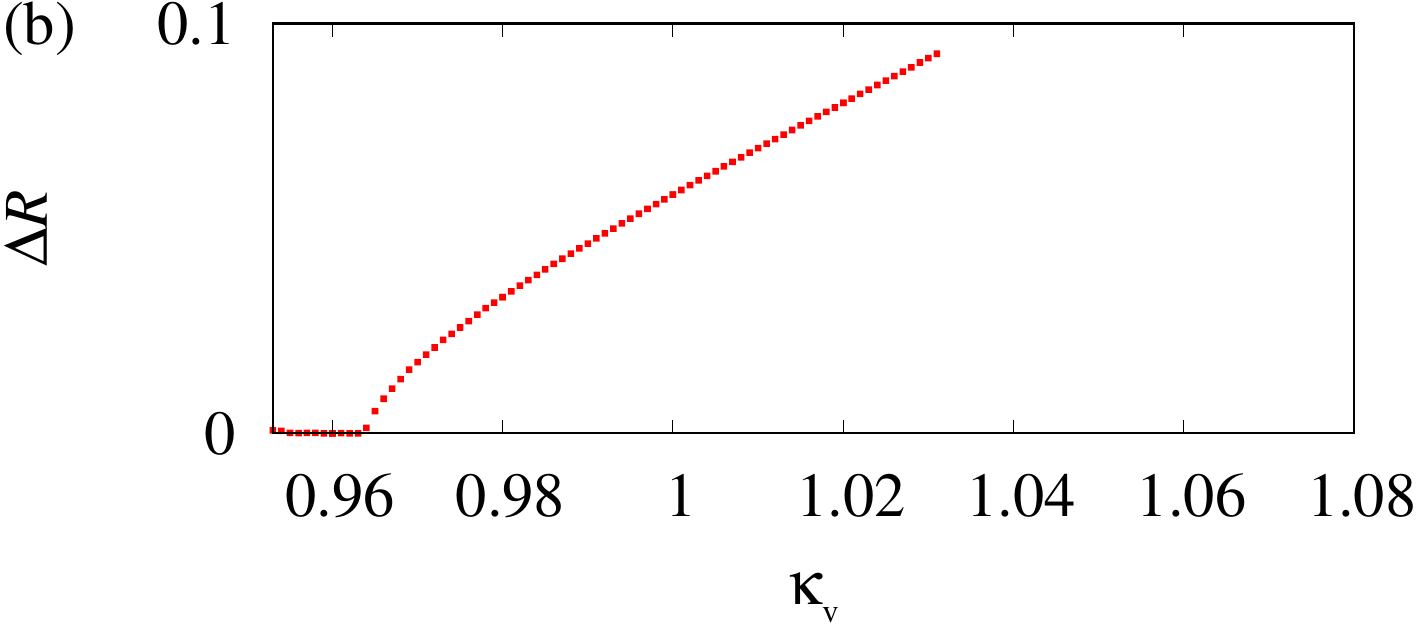}\\[2mm]
\includegraphics[width=0.45\textwidth]{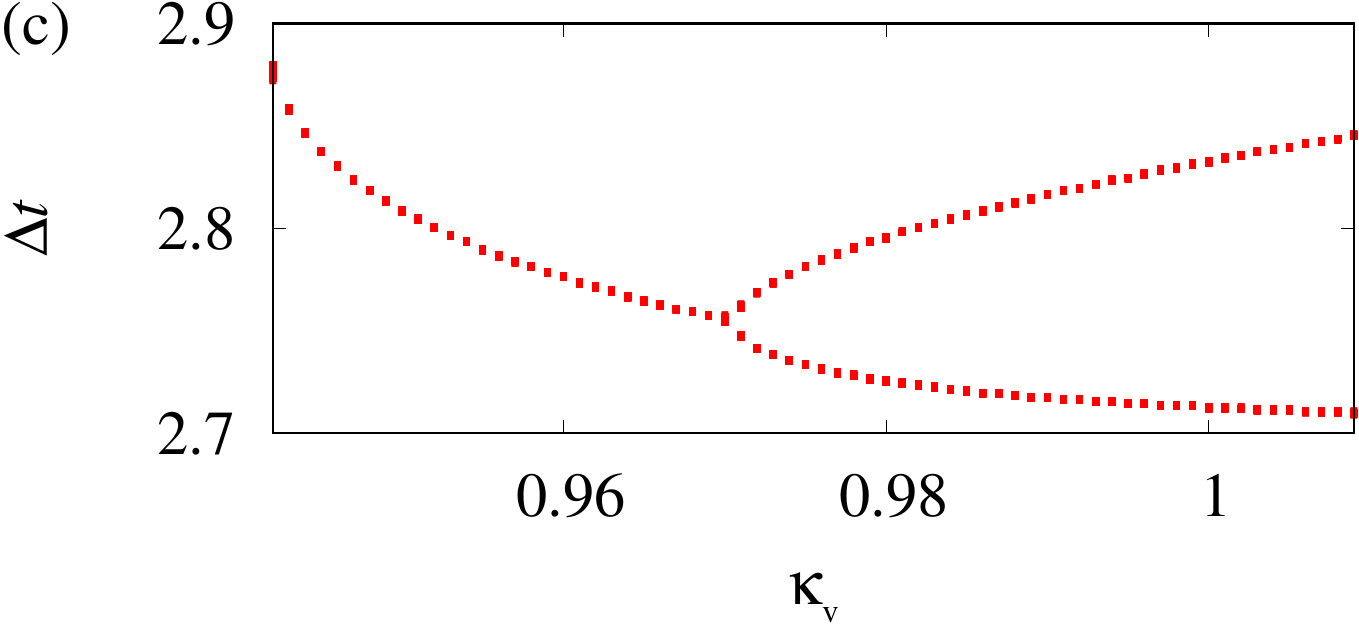}\hspace{8mm}
\includegraphics[width=0.45\textwidth]{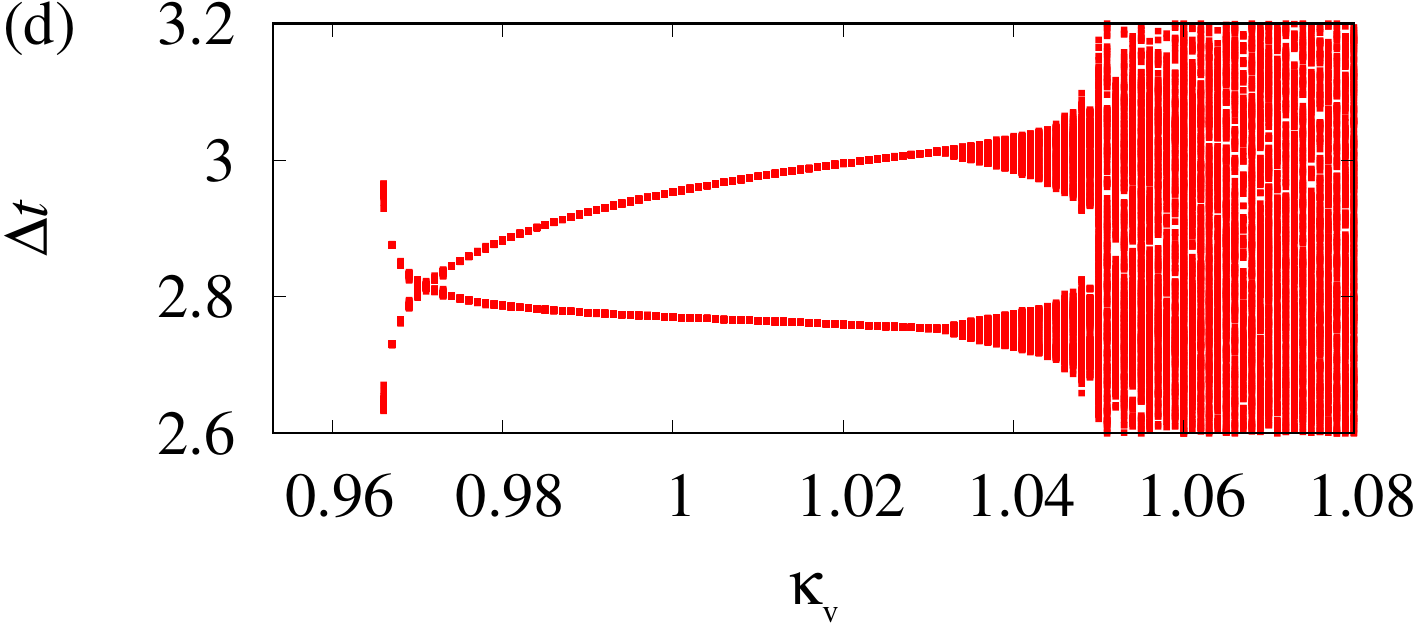}\\[2mm]
\includegraphics[width=0.45\textwidth]{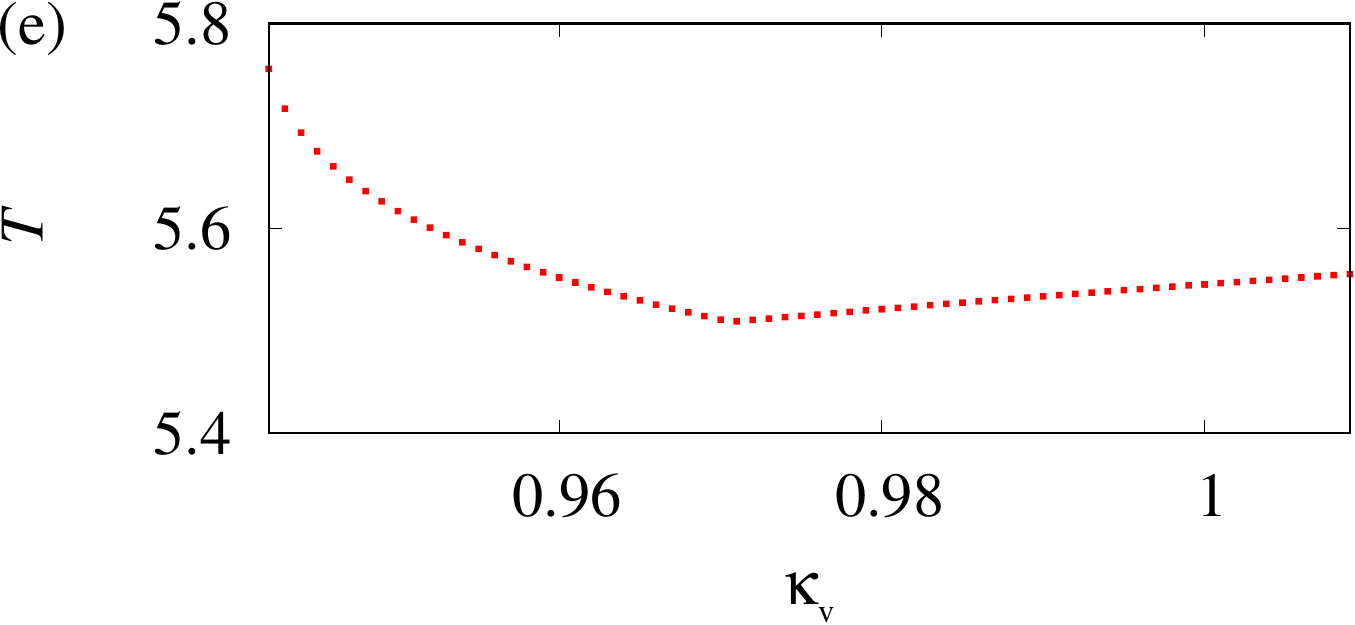}\hspace{8mm}
\includegraphics[width=0.45\textwidth]{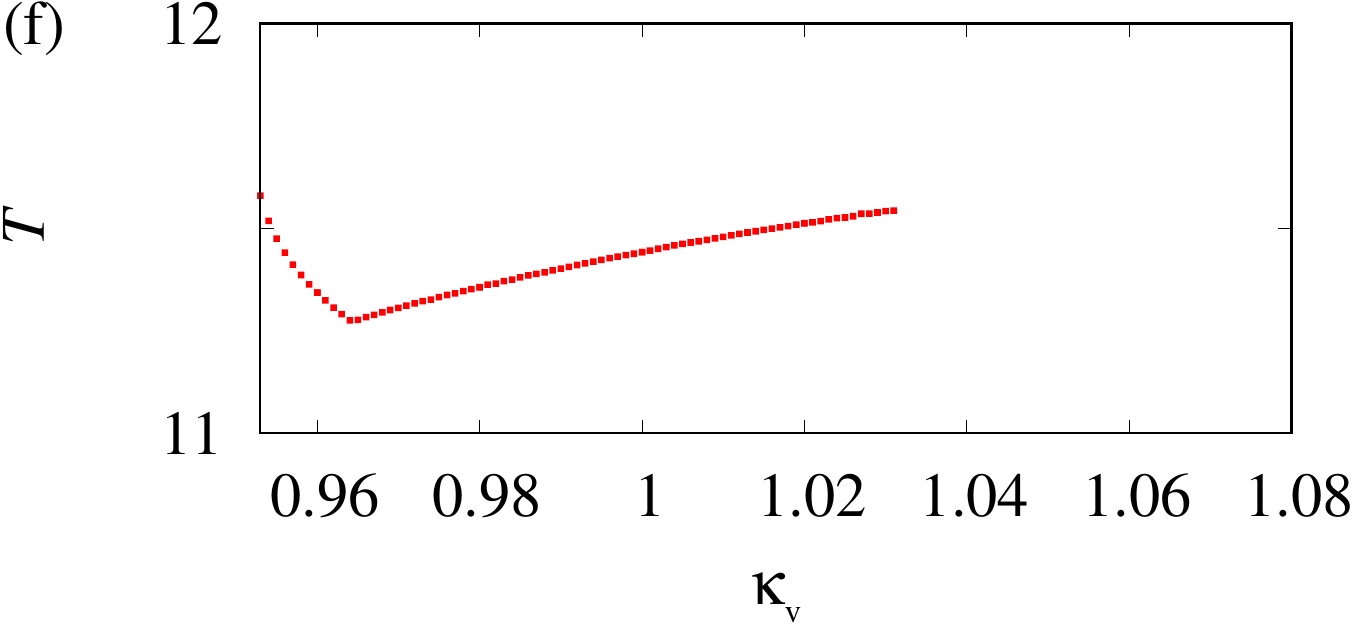}
\caption{Parameter sweeps of standing wave (left column)
and travelling wave (right column) solutions of Eq.~(\ref{Eq:MeanField}).
Three rows show the time variation of spatially averaged firing rate $\Delta R$,
the time intervals $\Delta t$ between the consecutive local maxima in the plot of $\langle R(t) \rangle$,
and the least period $T$.
Other parameters: $\eta_0 = 1$, $\gamma = 0.5$ and $\kappa_\mathrm{s} = 10$.}
\label{Fig:Scan}
\end{figure}

Next, by adiabatically changing the coupling strength $\kappa_\mathrm{v}$,
we investigated the stability intervals of these patterns.
(When moving from one value of $\kappa_\mathrm{v}$ to the next,
we added a small random perturbation to the final state to give the new initial state.)
To this end, for each value of $\kappa_\mathrm{v}$,
we calculated a trajectory of length $1000$ time units,
discarding the previous transient of the same length.
To characterize the dynamics of the obtained trajectory,
we used the spatially averaged firing rate
$$
\langle R(t) \rangle = \fr{1}{2\pi} \int_0^{2\pi} R(x,t) dx = \fr{1}{2\pi^2} \int_0^{2\pi} \Real u(x,t) dx
$$
and its time variation
$$
\Delta R = \max\limits_t \langle R(t) \rangle - \min\limits_t \langle R(t) \rangle.
$$
where $t$ runs over the length of the simulation.
In the case $\Delta R > 0$, i.e. when $\langle R(t) \rangle$ is non-constant,
we used the plot of $\langle R(t) \rangle$
to calculate the time intervals $\Delta t$ between consecutive local maxima of this function.
The resulting dependences of $\Delta R$ and $\Delta t$ on $\kappa_\mathrm{v}$
are shown in Fig.~\ref{Fig:Scan}. 
Moreover, for every time-periodic solution of Eq.~(\ref{Eq:MeanField}),
we also show its least period $T$, which was calculated by averaging
the time intervals between the consecutive local maxima of $\Real u(0,t)$.
(Note that for travelling waves we needed to double the above value,
since $\Real u(0,t)$ attains two maxima during the period.
Moreover, for travelling waves with $\kappa_\mathrm{v} > 0.964$
we also took into account that additional local maxima with $\Real\ u(0,t) < 1.5$
occur between the primary local maxima of $\Real\ u(0,t)$.)

\begin{figure}[h]
\includegraphics[width=0.45\textwidth]{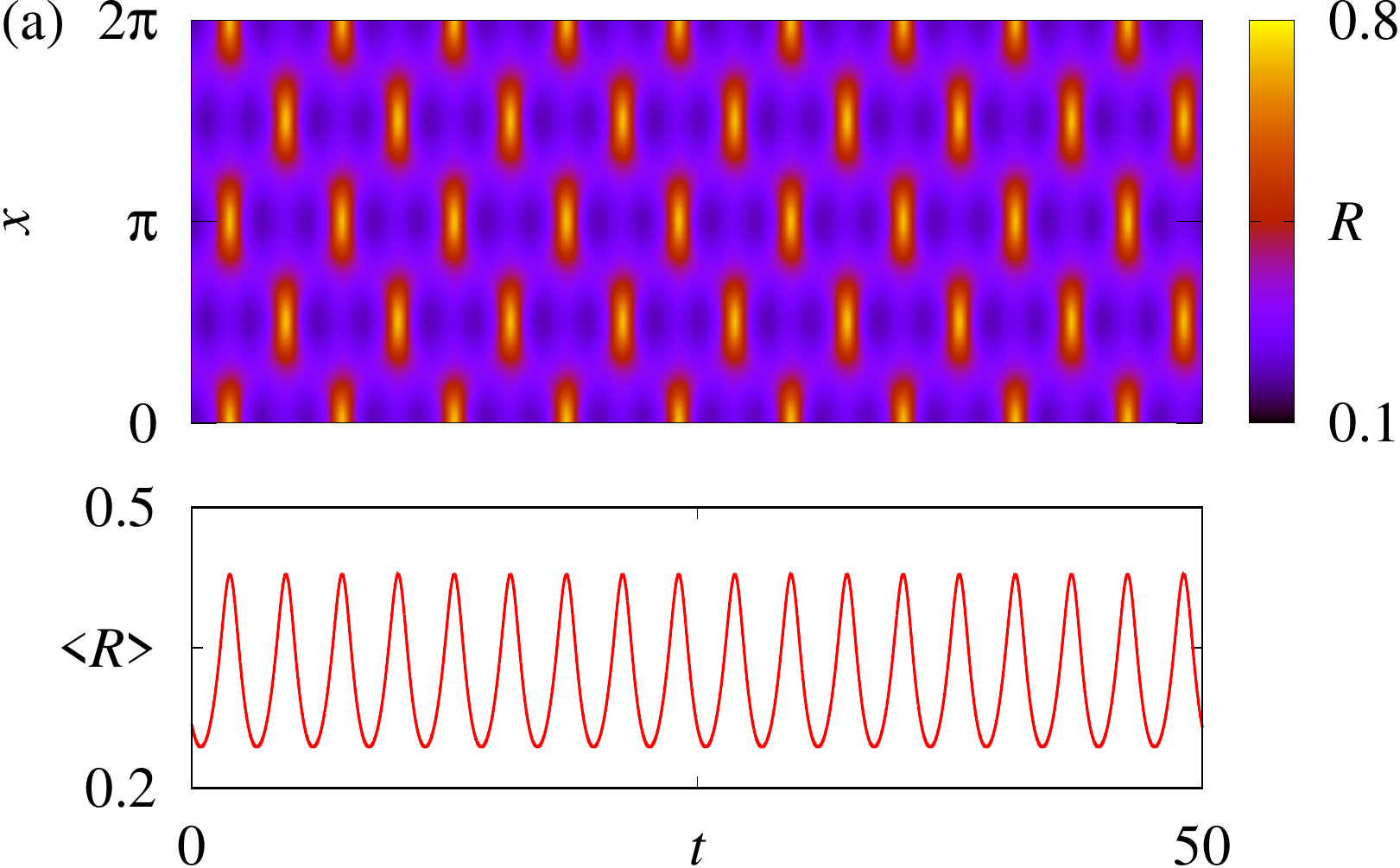}\hspace{5mm}
\includegraphics[width=0.45\textwidth]{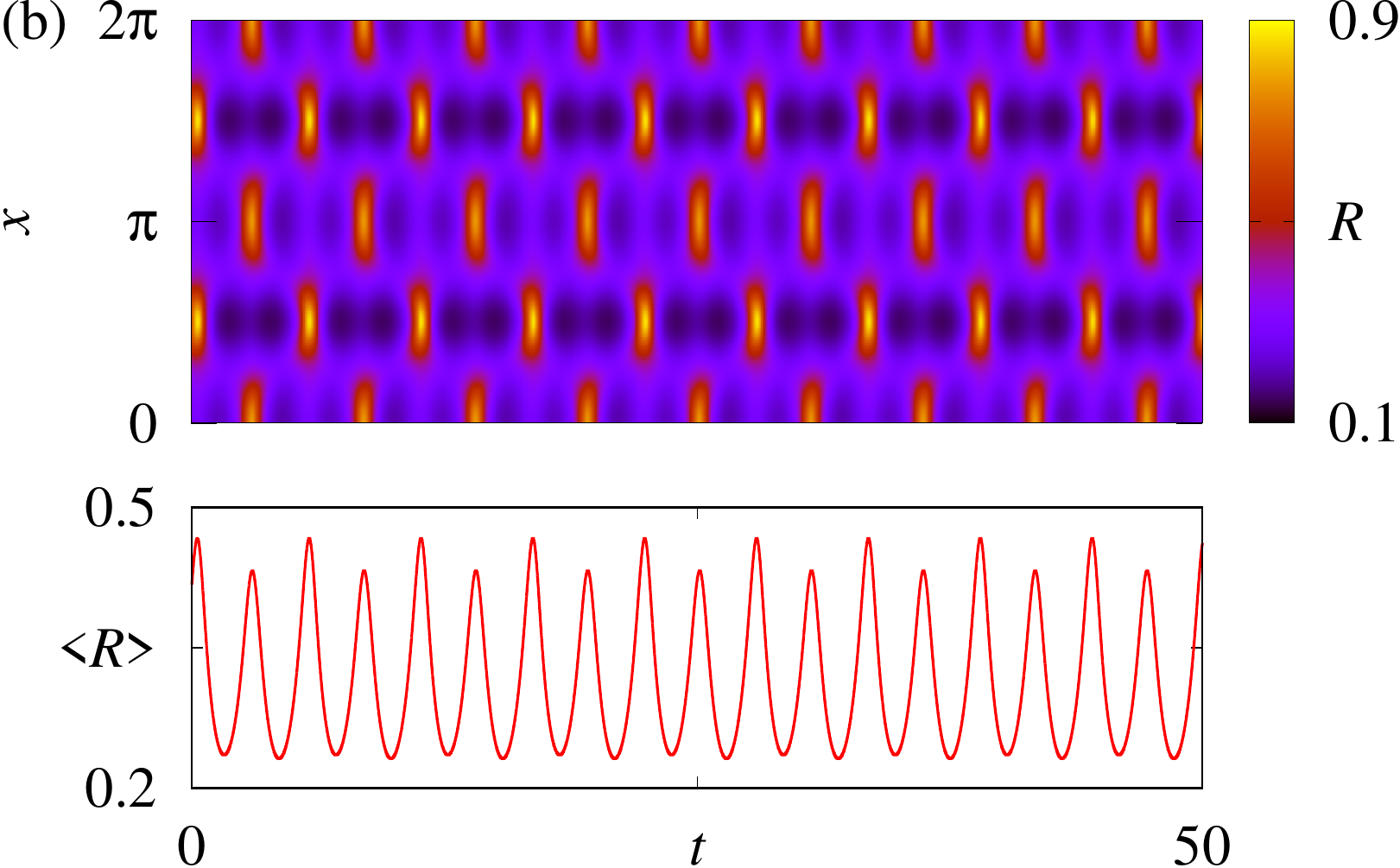}
\caption{Two types of standing wave solutions of Eq.~(\ref{Eq:MeanField})
for $\kappa_\mathrm{v} = 0.96$ (a) and $\kappa_\mathrm{v} = 0.99$ (b).
Other parameters: $\eta_0 = 1$, $\gamma = 0.5$ and $\kappa_\mathrm{s} = 10$.}
\label{Fig:SW}
\end{figure}

The diagrams in the left column of Fig.~\ref{Fig:Scan} indicate
the existence of two types of standing waves:
the waves with a single maximal value of $\langle R(t) \rangle$
and the waves with two alternating maximal values of $\langle R(t) \rangle$.
Typical examples of these patterns are shown in Fig.~\ref{Fig:SW}.

\begin{figure}[h]
\includegraphics[width=0.45\textwidth]{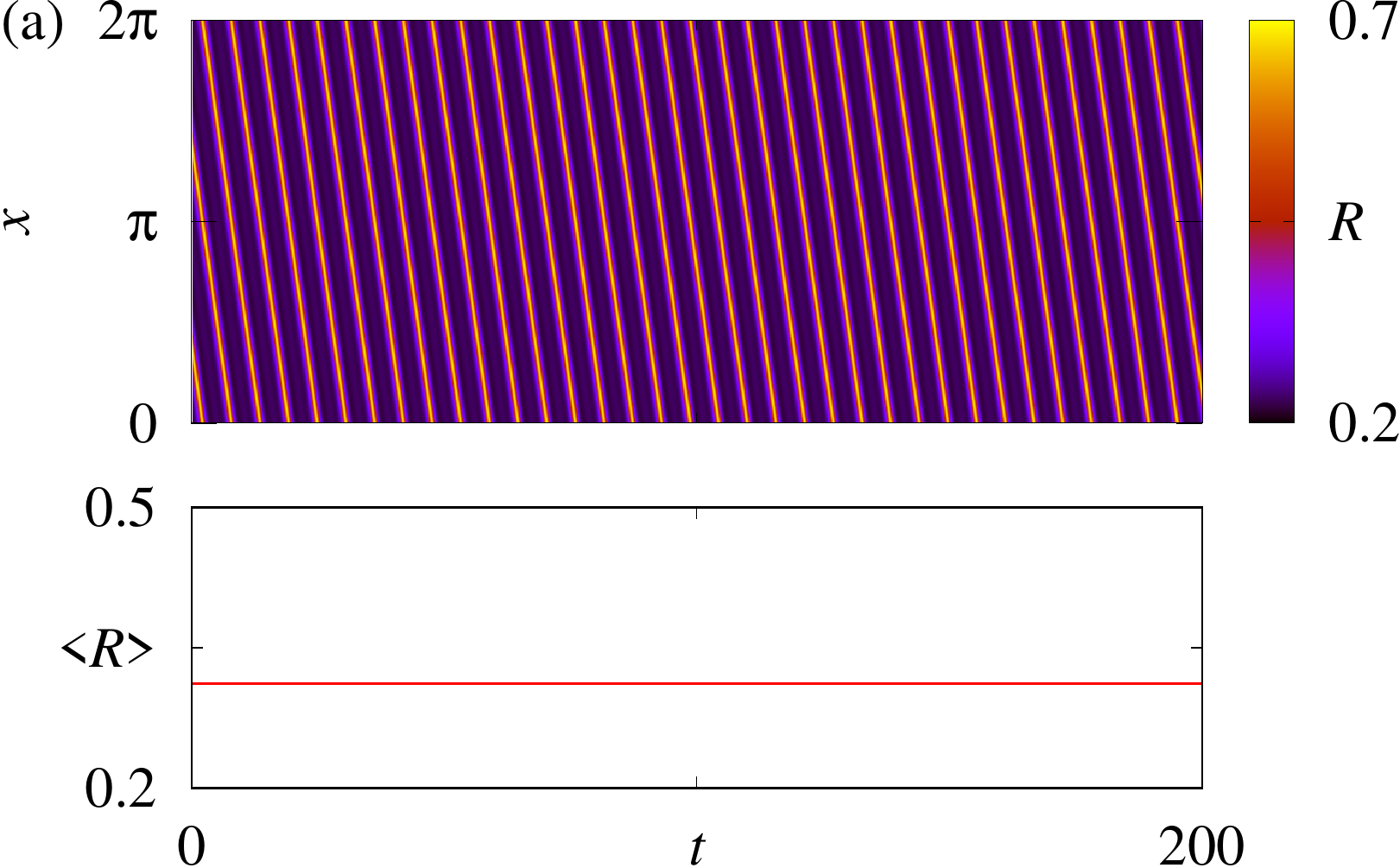}\hspace{5mm}
\includegraphics[width=0.45\textwidth]{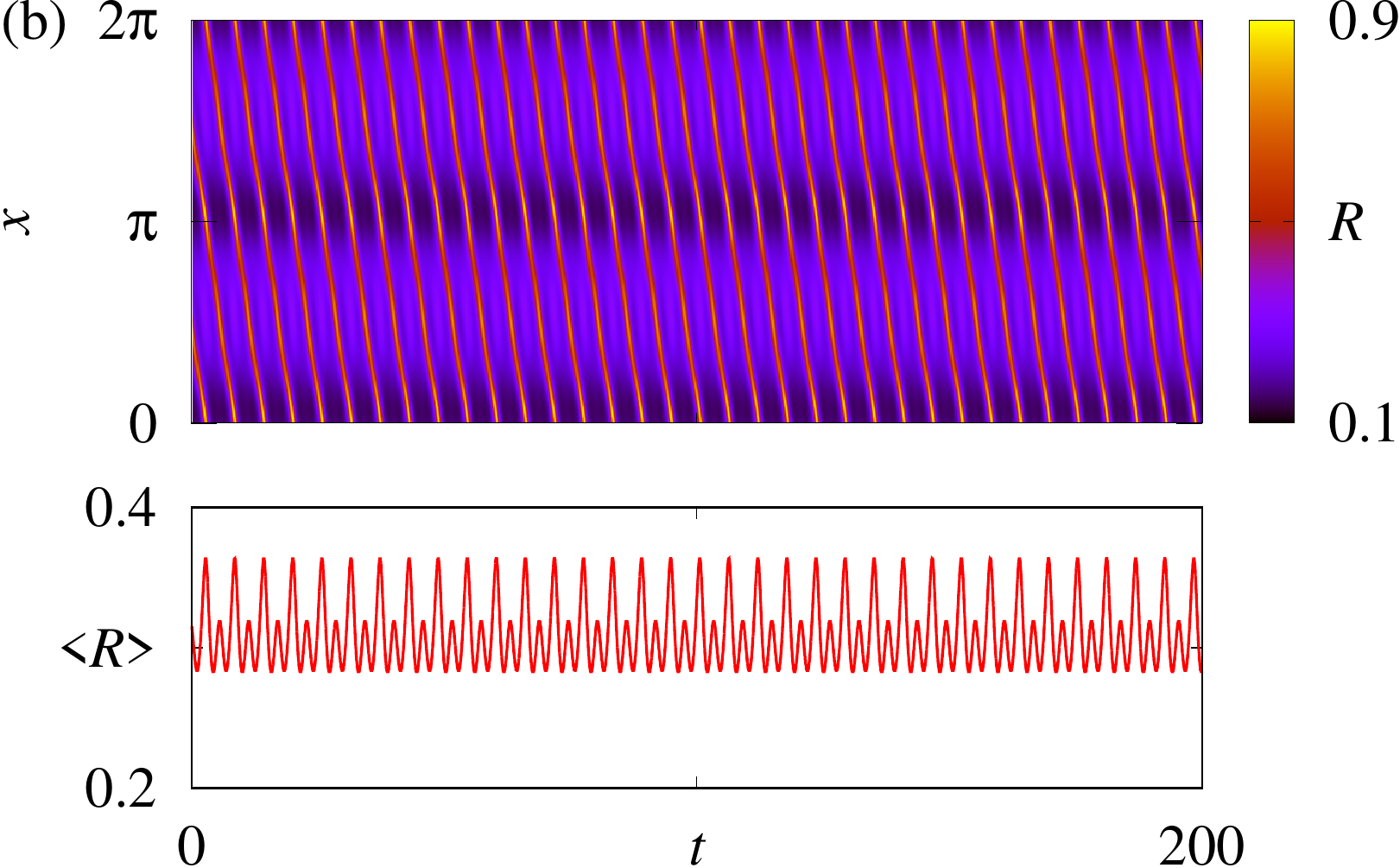}\\[4mm]
\includegraphics[width=0.45\textwidth]{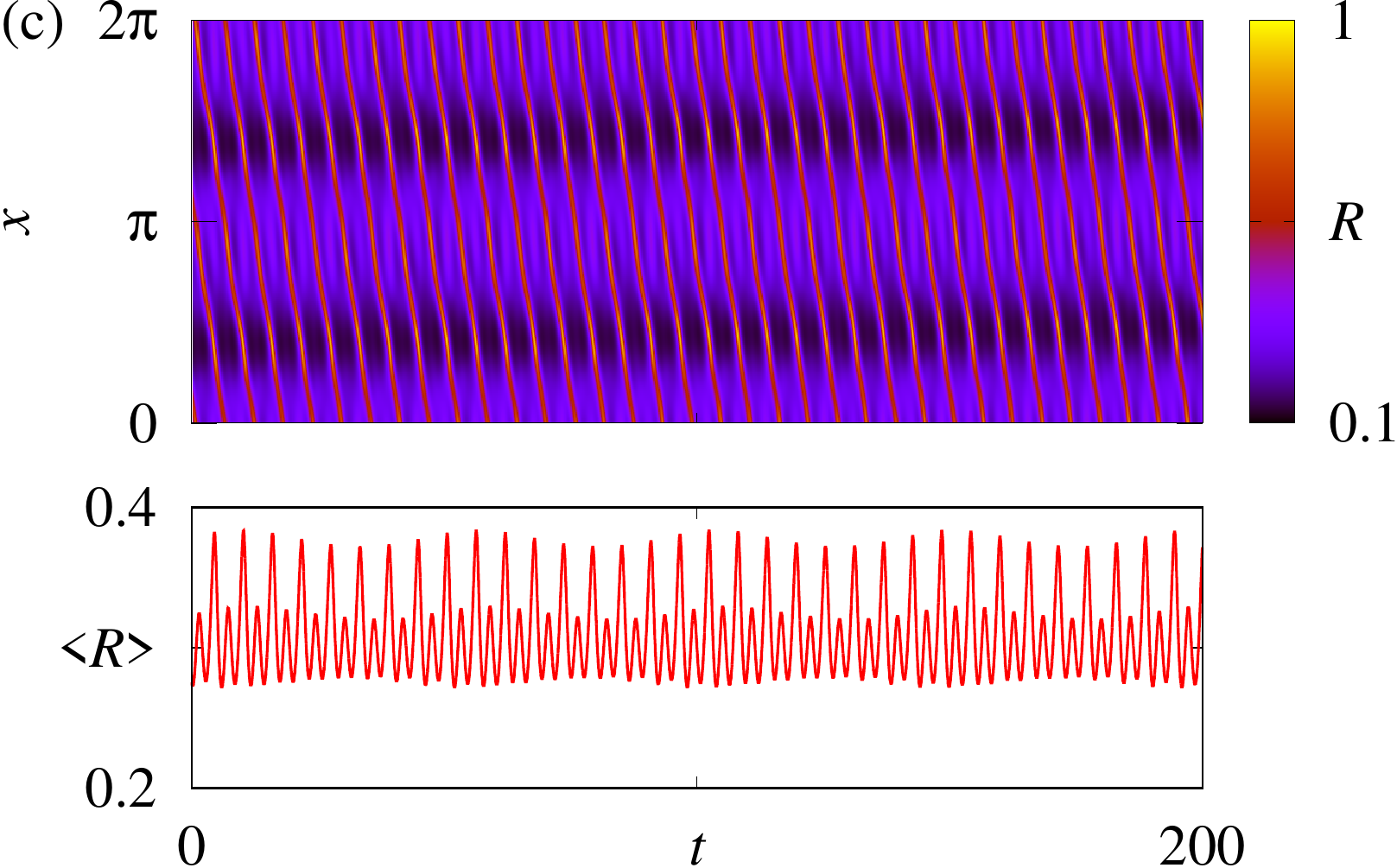}\hspace{5mm}
\includegraphics[width=0.45\textwidth]{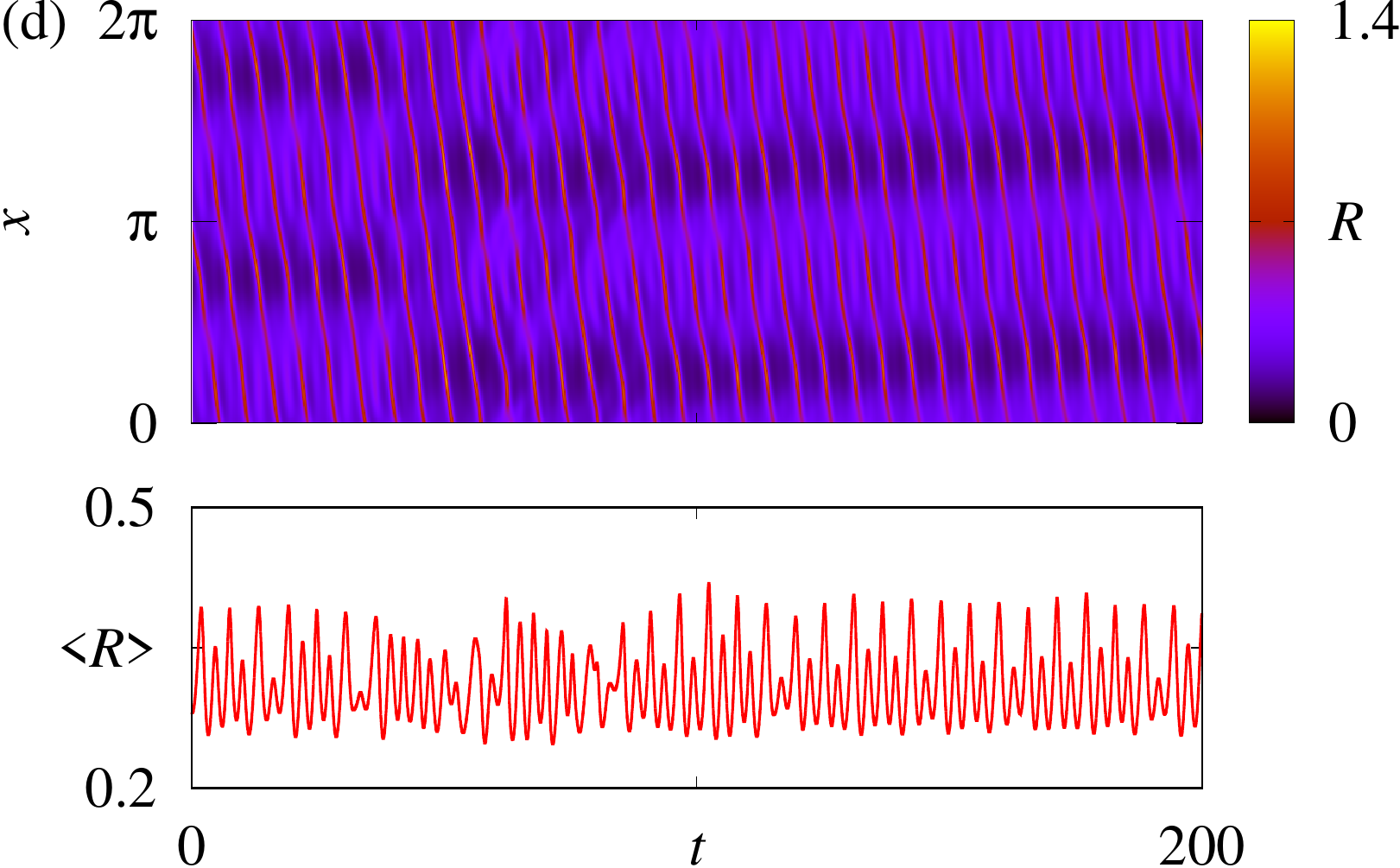}
\caption{Four types of travelling waves in Eq.~(\ref{Eq:MeanField})
for $\kappa_\mathrm{v} = 0.96$ (a), $\kappa_\mathrm{v} = 1.02$ (b),
$\kappa_\mathrm{v} = 1.04$ (c) and $\kappa_\mathrm{v} = 1.07$ (d).
Other parameters: $\eta_0 = 1$, $\gamma = 0.5$ and $\kappa_\mathrm{s} = 10$.
}
\label{Fig:TW}
\end{figure}

The diagrams in the right column of Fig.~\ref{Fig:Scan} have a more complicated structure.
There, we can distinguish four different types of travelling waves, see Fig.~\ref{Fig:TW}.
The left-most values of $\kappa_\mathrm{v}$ correspond
to a wave that moves rigidly (notice $\Delta R = 0$) with a constant speed (Fig.~\ref{Fig:TW}(a)).
For a larger value of $\kappa_\mathrm{v}$, we found a wave propagating
with a non-constant profile, which slowly drifts (Fig.~\ref{Fig:TW}(b)). 
This is a {\em lurching} wave. For a still larger value of $\kappa_\mathrm{v}$ the lurching
wave undergoes some form of quasiperiodic modulation (Fig.~\ref{Fig:TW}(c)).
Finally, for the right-most values of $\kappa_\mathrm{v}$,
we observe a wave moving above a turbulent background (Fig.~\ref{Fig:TW}(d)).

\subsection{Case $\kappa_\mathrm{s} = 20$}

\begin{figure}[h]
\includegraphics[width=0.45\textwidth]{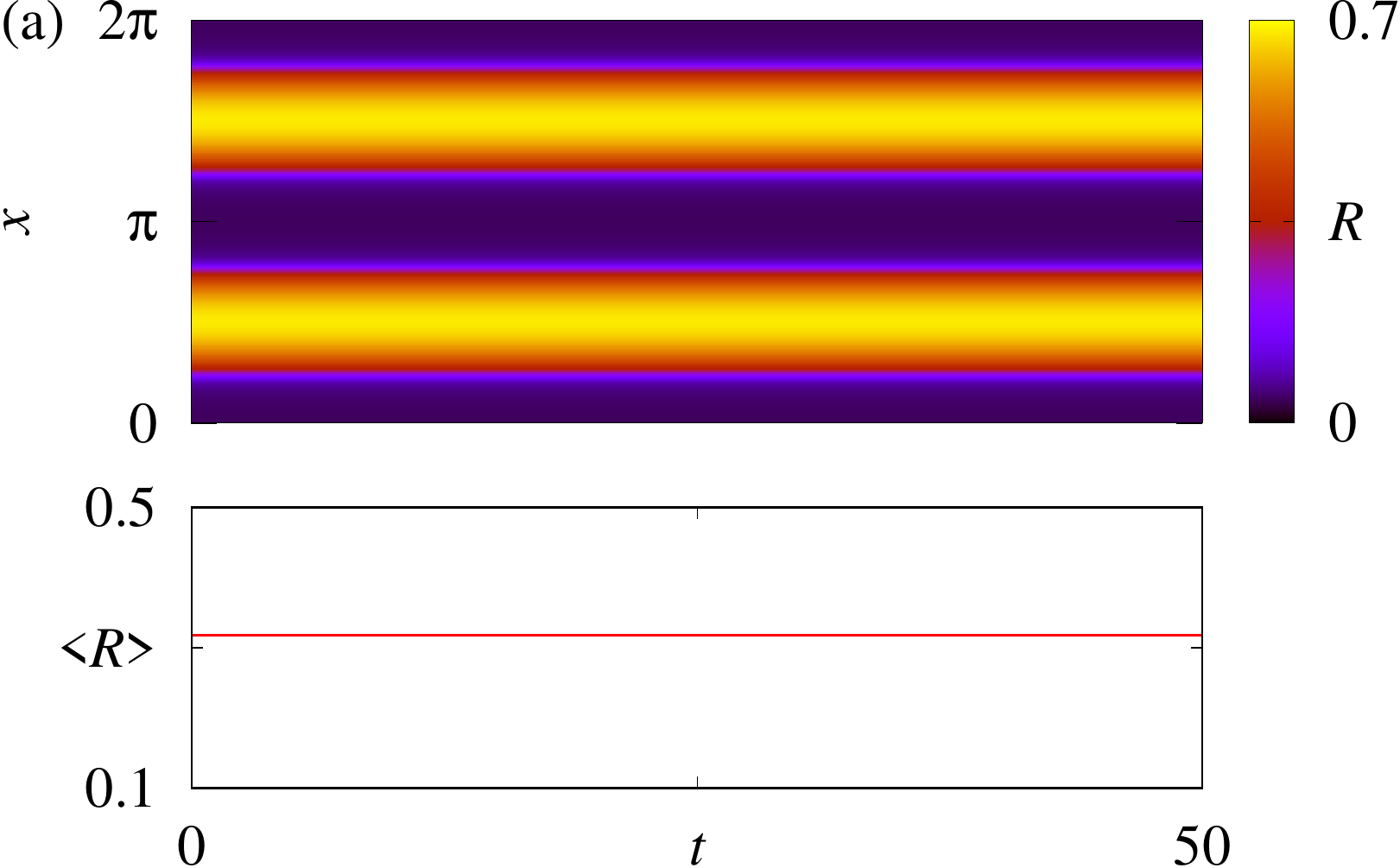}\hspace{5mm}
\includegraphics[width=0.45\textwidth]{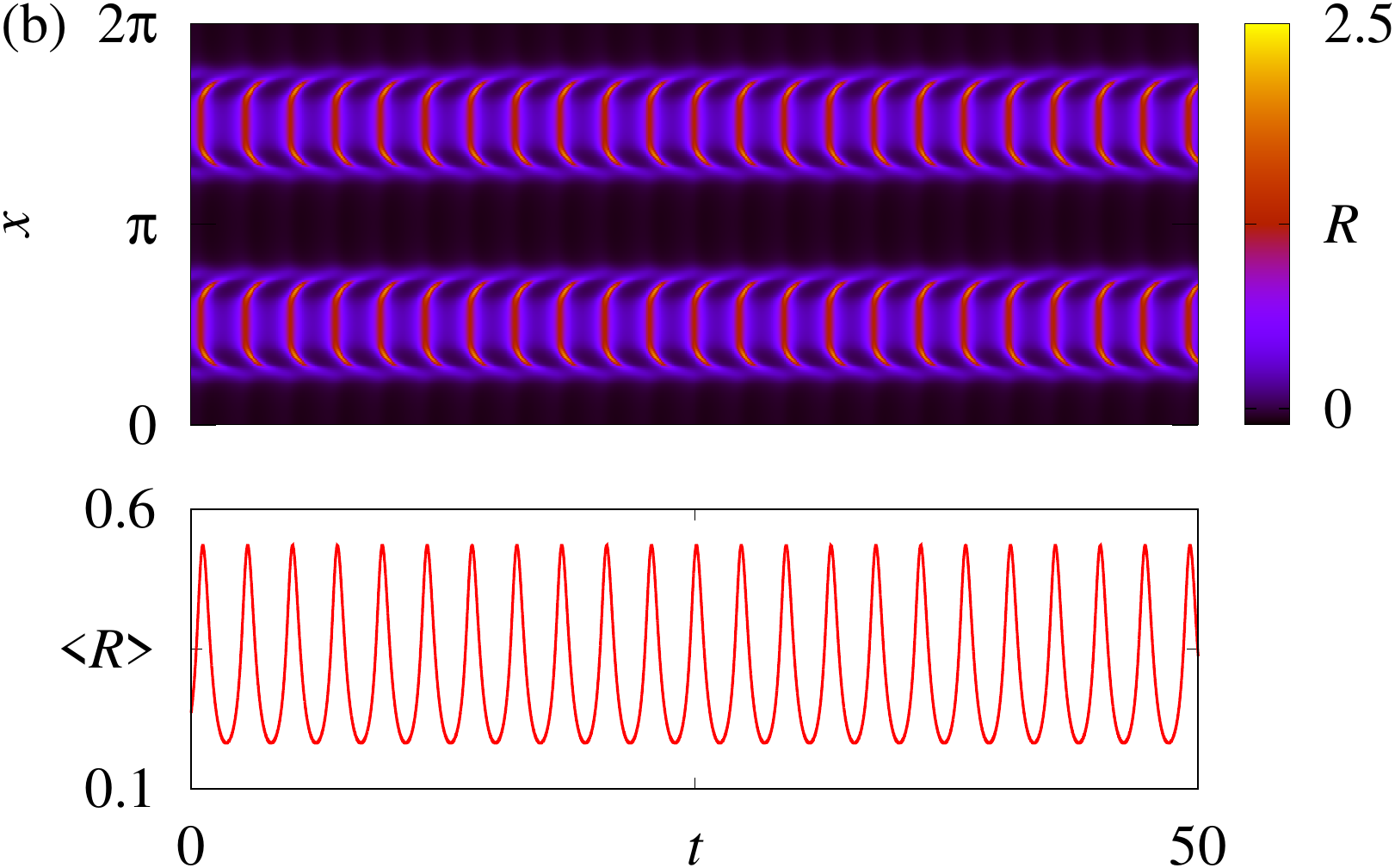}\\[4mm]
\includegraphics[width=0.45\textwidth]{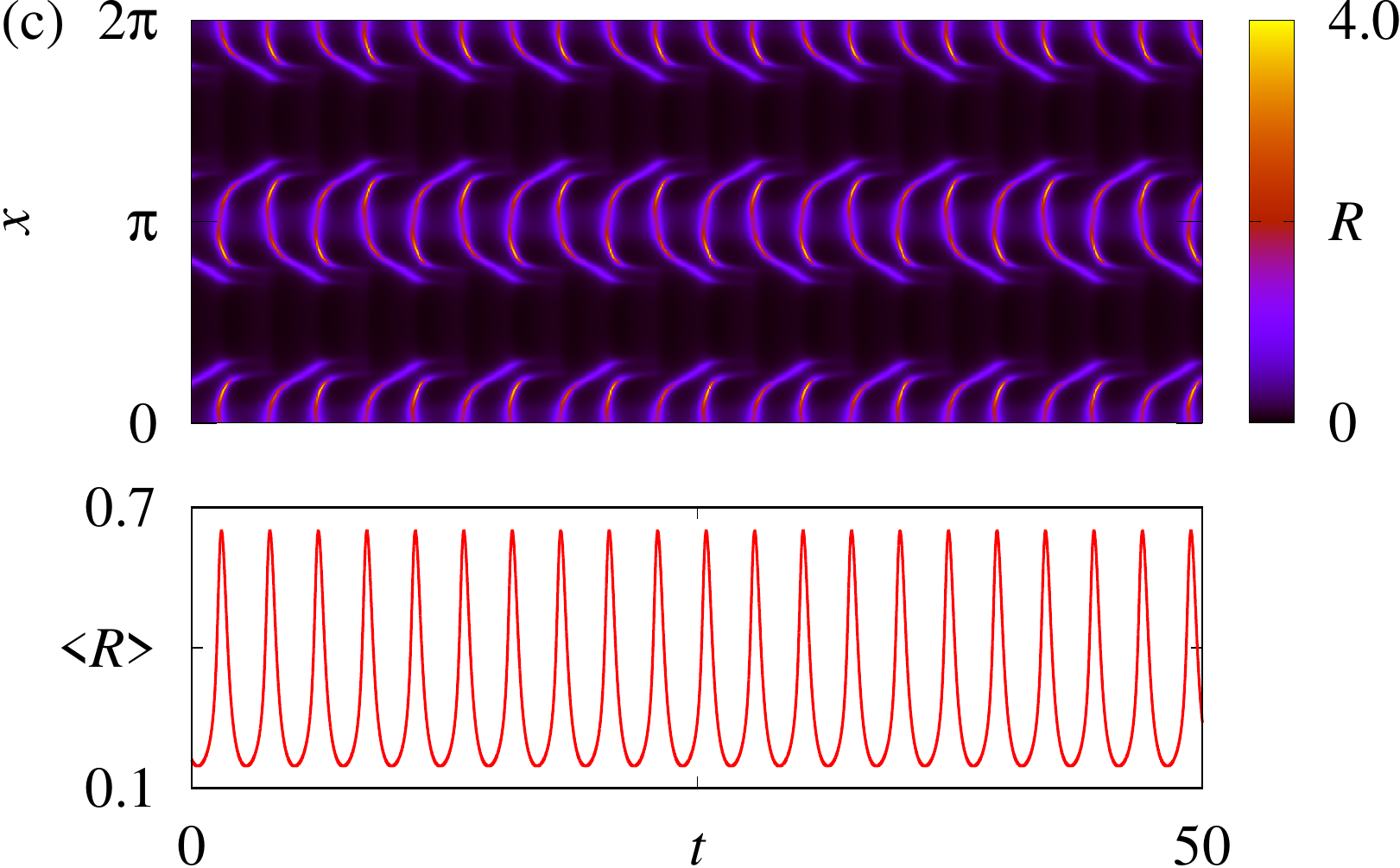}\hspace{5mm}
\includegraphics[width=0.45\textwidth]{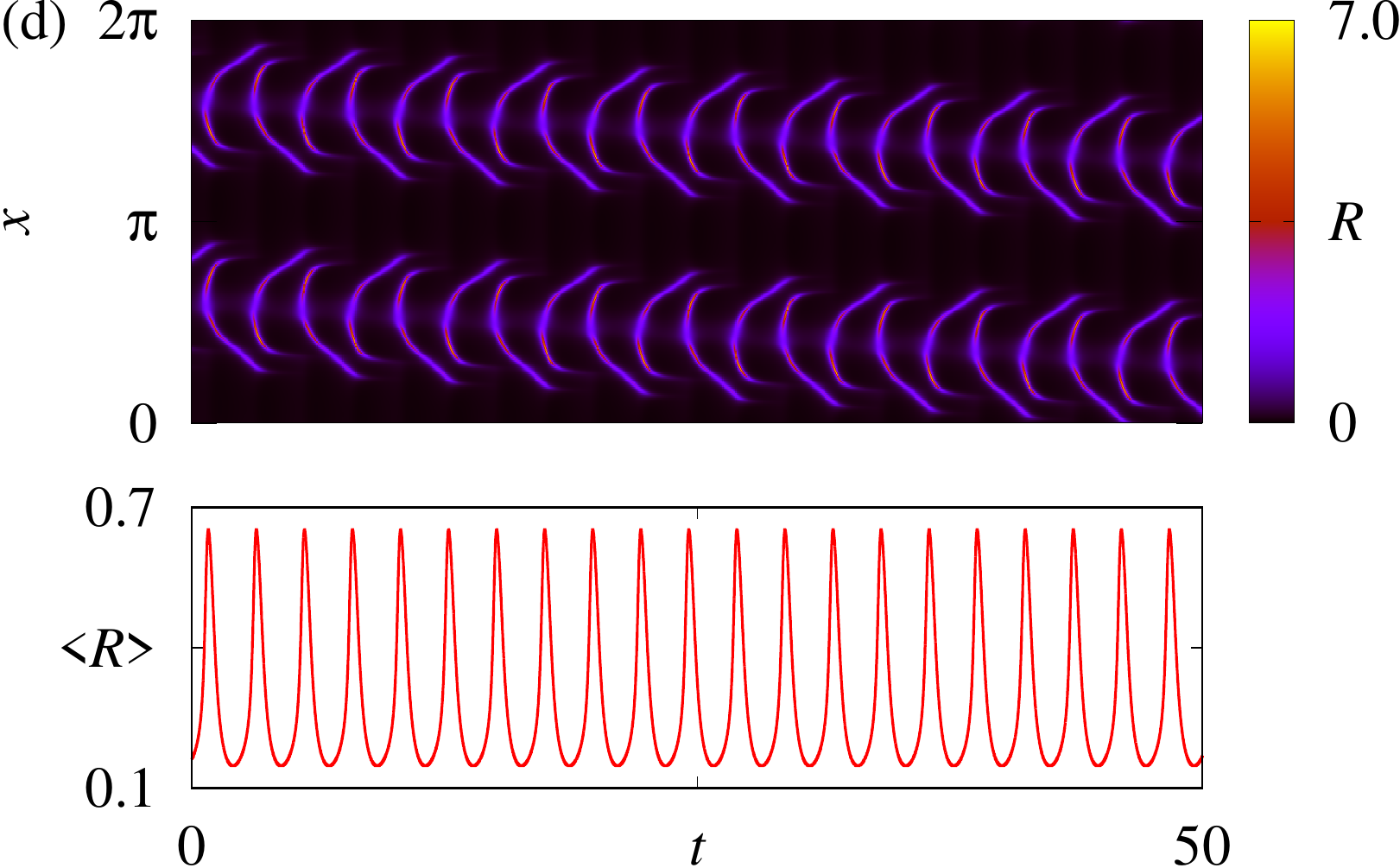}\\[4mm]
\includegraphics[width=0.45\textwidth]{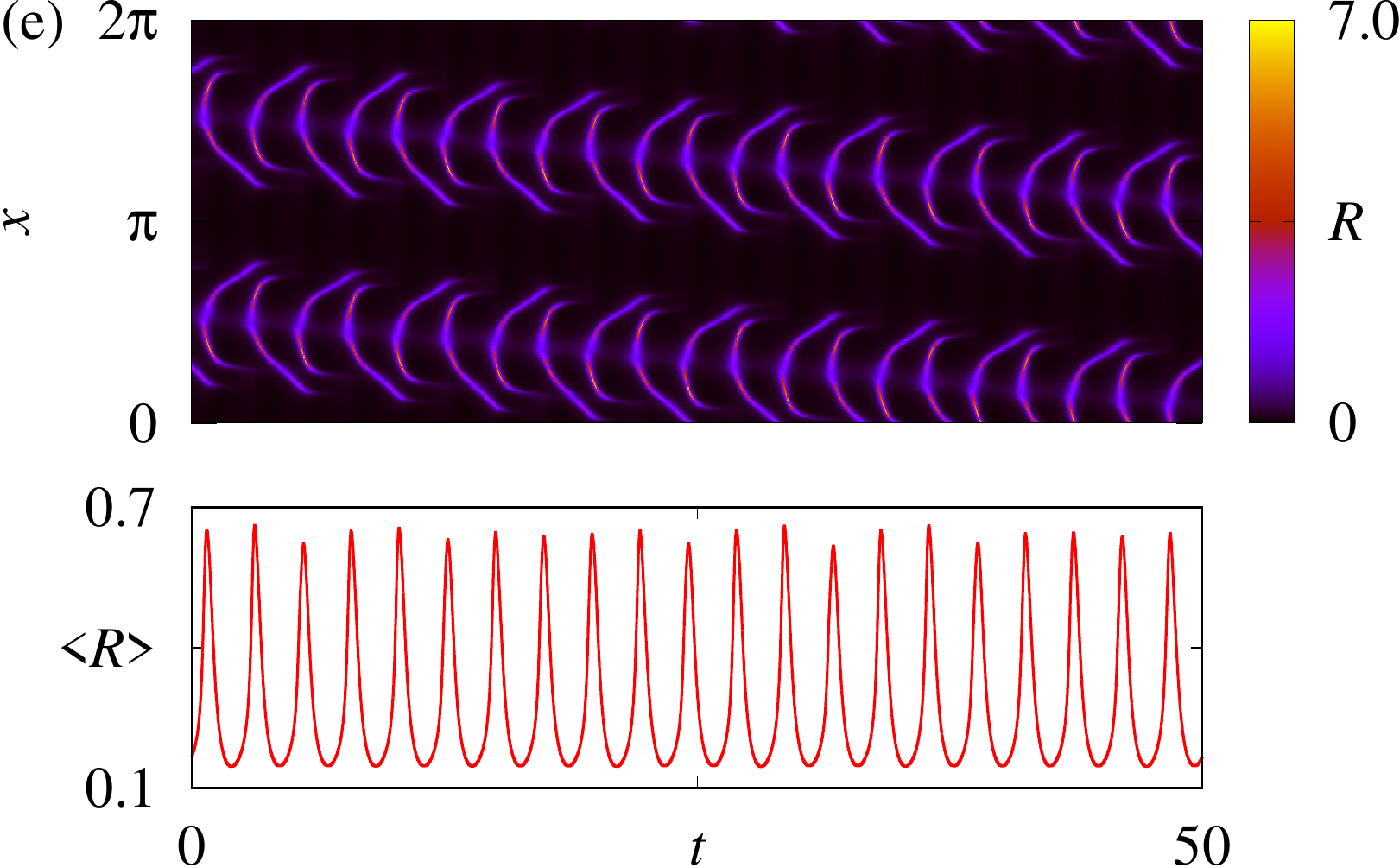}\hspace{5mm}
\includegraphics[width=0.45\textwidth]{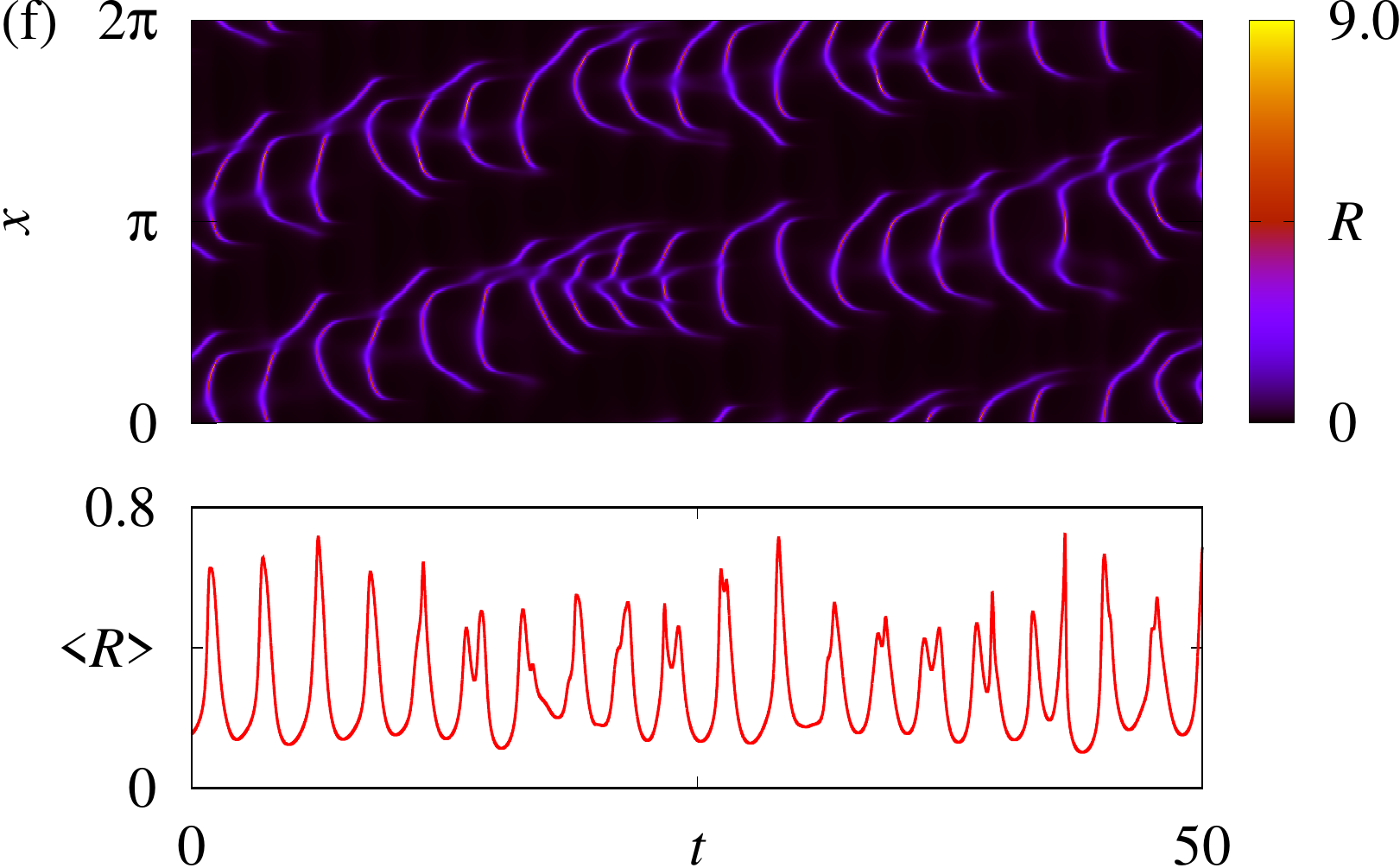}
\caption{Six types of solutions of Eq.~(\ref{Eq:MeanField})
for $\kappa_\mathrm{v} = 0.8$ (a), $\kappa_\mathrm{v} = 1$ (b),
$\kappa_\mathrm{v} = 1.3$ (c), $\kappa_\mathrm{v} = 1.6$ (d),
$\kappa_\mathrm{v} = 1.65$ (e) and $\kappa_\mathrm{v} = 1.8$ (f).
Other parameters: $\eta_0 = 1$, $\gamma = 0.5$ and $\kappa_\mathrm{s} = 20$.
}
\label{Fig:Ks20_00}
\end{figure}

\begin{figure}[h]
\includegraphics[width=0.45\textwidth]{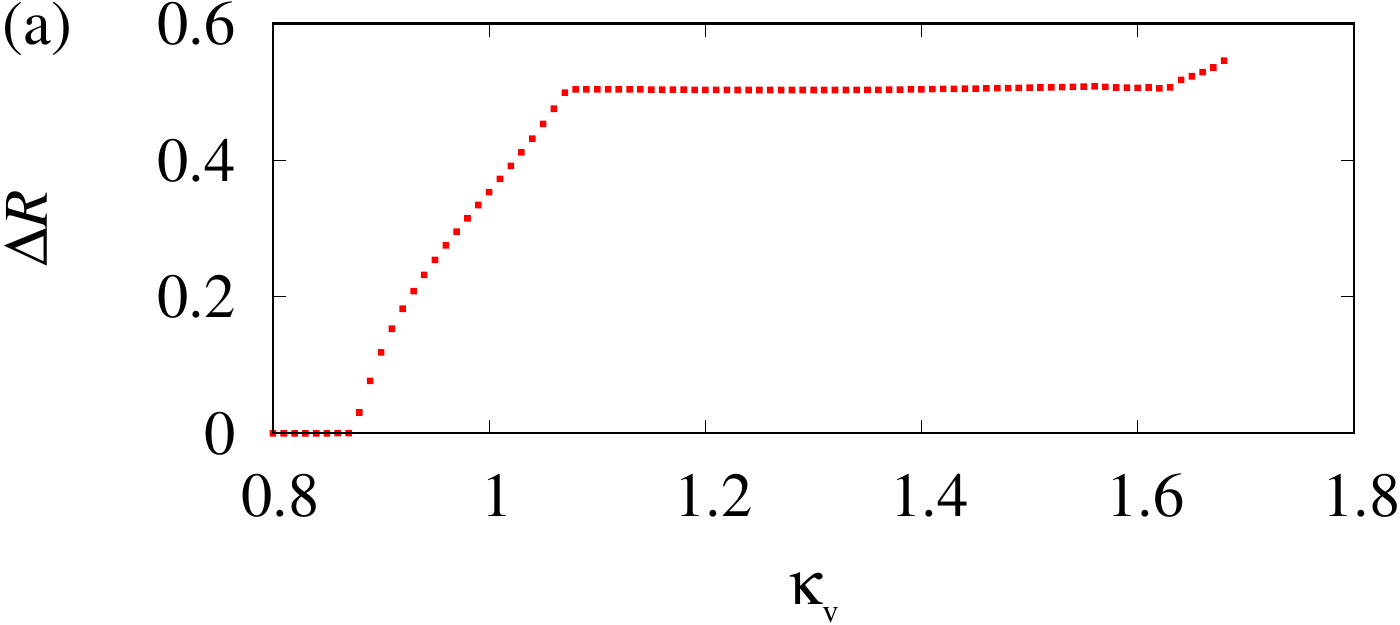}\hspace{8mm}
\includegraphics[width=0.45\textwidth]{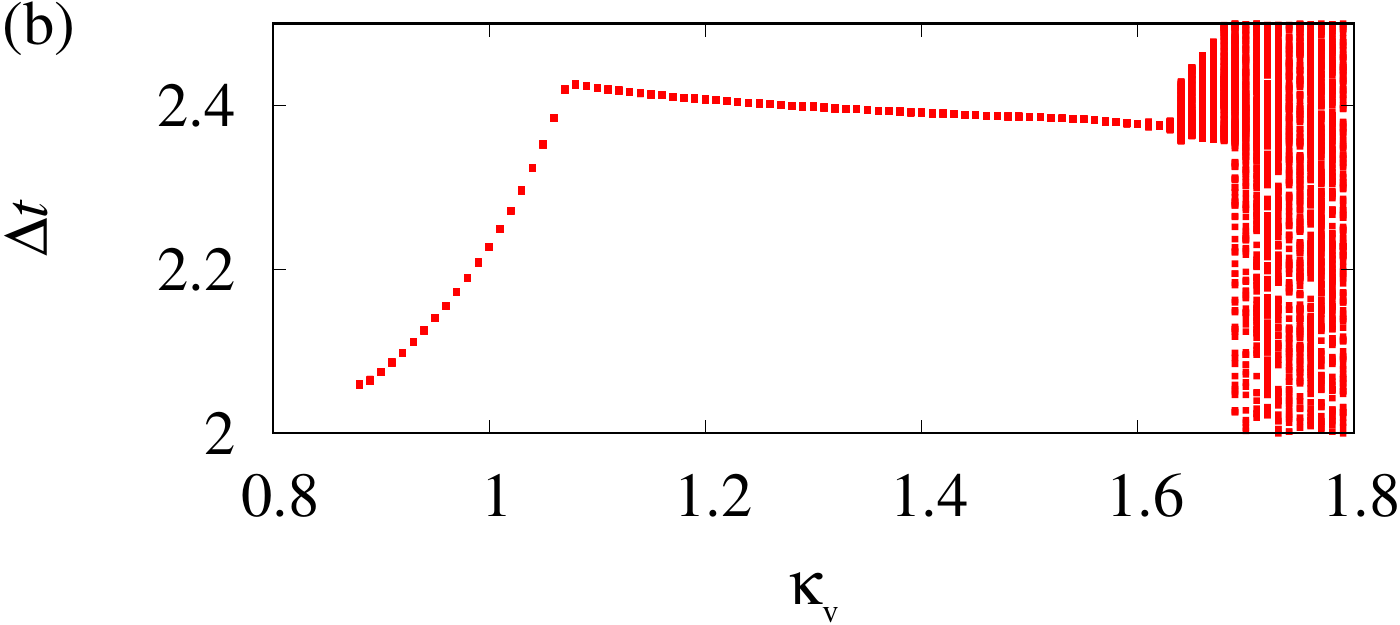}
\caption{A parameter sweep for patterned solutions shown in Fig.~\ref{Fig:Ks20_00}.
The panels (a) and (b) show the time variation of spatially averaged firing rate $\Delta R$
and the time intervals $\Delta t$ between the consecutive local maxima in the plot of $\langle R(t) \rangle$, respectively.
Other parameters: $\eta_0 = 1$, $\gamma = 0.5$ and $\kappa_\mathrm{s} = 20$.}
\label{Fig:Scan:Ks20_00}
\end{figure}

Keeping $\eta_0 = 1$, $\gamma = 0.5$ and setting $\kappa_\mathrm{s} = 20$
we obtain several new types of solution for different values of $\kappa_\mathrm{v}$;
see Fig.~\ref{Fig:Ks20_00}.
We see ``two-bump'' stationary solutions, Fig.~\ref{Fig:Ks20_00}(a),
for which there are two disjoint regions where neurons are firing
at a significant rate~\cite{laitro02,laitro03}.
We also see standing waves (or ``breathing'' two-bump solutions, similar to those
seen in~\cite{lai15,byravi19,schavi20}),
Fig.~\ref{Fig:Ks20_00}(b), more complex standing wave solutions,
Fig.~\ref{Fig:Ks20_00}(c), lurching waves, Fig.~\ref{Fig:Ks20_00}(d),
and waves that travel but are not periodic
in a uniformly-travelling coordinate frame, Fig.~\ref{Fig:Ks20_00}(e,f).
Quasi-statically sweeping through $\kappa_\mathrm{v}$ we obtain 
Fig.~\ref{Fig:Scan:Ks20_00}.

In the rest of the paper we analyse the types of solution shown, and provide a more rigorous
explanation for the results observed above.

\section{Analytical methods}
\label{sec:bif}

The kaleidoscopic set of spatiotemporal patterns found in Eq.~(\ref{Eq:MeanField})
for various system parameters can be logically explained
by a bifurcation analysis of this equation, focusing
on its equilibrium and periodic solutions.
We describe two ways to do this.
Our first approach relies on a discretized version
of the integro-differential equation~(\ref{Eq:MeanField})
and uses a standard implementation in Matlab of pseudo-arclength continuation for the bifurcation analysis
of finite-dimensional dynamical systems~\cite{lai14b}.
In the second approach, we derive a set of integral self-consistency equations
for different types of solutions of Eq.~(\ref{Eq:MeanField})
and carry out their finite-dimensional reduction by Galerkin method.

\subsection{Uniform states}
\label{Sec:Uniform}

We start with the simplest form of solution, a uniform state.
Recall that these are constant solutions of Eq.~(\ref{Eq:MeanField}),
which do not depend on either $x$ or $t$.
On the other hand, these states
will not be steady states of system~(\ref{Eq:Nonlocal}), as that describes
a finite network of heterogeneous spiking neurons. However, in system~(\ref{Eq:Nonlocal}),
such a state would have no macroscopic spatial structure nor significant temporal variations.
 
\subsubsection{Existence}

Each uniform state of Eq.~(\ref{Eq:MeanField}) corresponds
to a solution $u\in\mathbb{C}$ of the equation
\begin{equation}
\gamma - \kappa_\mathrm{v} u + i \left[ \eta_0 + \kappa_\mathrm{v} \mathcal{K}_\mathrm{v} \Imag(u) + \fr{\kappa_\mathrm{s}}{\pi} \mathcal{K}_\mathrm{s} \Real(u) - u^2 \right] = 0.
\label{Eq:FP}
\end{equation}
In the next proposition, we show that all physically meaningful solutions
of Eq.~(\ref{Eq:FP}) lie on a certain one-parameter manifold.

\begin{proposition}
All constant solutions of Eq.~(\ref{Eq:FP}) satisfying $\Real(u) \ge 0$ lie
on the manifold defined by formulas
\begin{eqnarray}
u &=& \fr{1}{2} f_-(F,\kappa_\mathrm{v},\gamma) + \fr{i}{2} \left( \kappa_\mathrm{v} - f_+(F,\kappa_\mathrm{v},\gamma) \right),
\label{FP:u}\\[2mm]
\eta_0 &=& F - \pi \kappa_\mathrm{v} \hat{W}_{\mathrm{v},0}
\left( \kappa_\mathrm{v} - f_+(F,\kappa_\mathrm{v},\gamma) \right) -  \kappa_\mathrm{s} \hat{W}_{\mathrm{s},0} f_-(F,\kappa_\mathrm{v},\gamma),
\label{FP:eta0}
\end{eqnarray}
where
$$
f_\pm(F,\kappa_\mathrm{v},\gamma) = \sqrt{ \fr{\sqrt{ ( \kappa_\mathrm{v}^2 - 4 F )^2 + 16 \gamma^2 }
\pm ( \kappa_\mathrm{v}^2 - 4 F )}{2} },
$$
the coefficients $\hat{W}_{\mathrm{v},0}$ and $\hat{W}_{\mathrm{s},0}$
are defined by the formula~(\ref{Fourier:W}) with $m=0$,
and $F\in\mathbb{R}$ is a free parameter.
\label{Proposition:FP}
\end{proposition}

{\bf Proof:}
Let us rewrite Eq.~(\ref{Eq:FP}) in the form
\begin{equation}
\gamma + i F - \kappa_\mathrm{v} u - i u^2 = 0,
\label{Eq:FP_}
\end{equation}
where
\begin{equation}
F = \eta_0 + \kappa_\mathrm{v} \mathcal{K}_\mathrm{v} \Imag(u) + \fr{\kappa_\mathrm{s}}{\pi} \mathcal{K}_\mathrm{s} \Real(u).
\label{Eq:F:Uniform}
\end{equation}
For a given $F$, Eq.~(\ref{Eq:FP_}) has one and only one solution
satisfying $\Real(u) \ge 0$, see Proposition~\ref{Proposition:FP:General}.
It is determined by the formula~(\ref{FP:u}).
Inserting this result into Eq.~(\ref{Eq:F:Uniform}) we obtain
\begin{eqnarray*}
F &=& \eta_0 + \pi \kappa_\mathrm{v} \hat{W}_{\mathrm{v},0}
\left( \kappa_\mathrm{v} -
\sqrt{ \fr{\sqrt{ ( \kappa_\mathrm{v}^2 - 4 F )^2 + 16 \gamma^2 }
+ \kappa_\mathrm{v}^2 - 4 F}{2} } \right) \\[2mm]
&+& \kappa_\mathrm{s} \hat{W}_{\mathrm{s},0} \sqrt{ \fr{\sqrt{ ( \kappa_\mathrm{v}^2 - 4 F )^2 + 16 \gamma^2 }
- \kappa_\mathrm{v}^2 + 4 F}{2} },
\end{eqnarray*}
what is an equivalent form of~(\ref{FP:eta0}).~\qed

Note that Proposition~\ref{Proposition:FP} allows us
to explicitly express the dependence
of the constant solution $u$ on the parameter $\eta_0$
(albeit, in a parametric form).
The same applies to the dependence of $u$ on $\kappa_\mathrm{s}$,
if $\hat{W}_{\mathrm{s},0}\ne 0$.

\subsubsection{Linear stability}
\label{Sec:US:LS}

Suppose that $u_0\in\mathbb{C}$ is a constant solution of Eq.~(\ref{Eq:FP}).
To analyze its stability, we insert the ansatz $u(x,t) = u_0 + v(x,t)$ into Eq.~(\ref{Eq:FP})
and linearize the resulting equation with respect to small perturbations $v(x,t)$.
This yields a linear integro-differential equation
\begin{equation}
\pf{v}{t} = \mu v + \fr{\kappa_\mathrm{v}}{2} \mathcal{K}_\mathrm{v} ( v - \overline{v} ) + \fr{i \kappa_\mathrm{s}}{2\pi} \mathcal{K}_\mathrm{s} ( v + \overline{v} ),
\label{Eq:Linear}
\end{equation}
where
\begin{equation}
\mu = - \kappa_\mathrm{v} - 2 i u_0.
\label{Def:mu}
\end{equation}

\begin{remark}
If $\Real(u_0)\ge 0$ and hence $u_0$ is determined by Proposition~\ref{Proposition:FP}, then $\Real(\mu) < 0$. Indeed, this inequality is easy to verify
by inserting the $u_0$ defined by formula~(\ref{FP:u}) into~(\ref{Def:mu}).
\label{Remark:SignOfMu}
\end{remark}

To investigate the growth or decay of different spatial modes, we insert the ansatz
$$
v(x,t) = v_+(x) e^{\lambda t} + \overline{v}_-(x) e^{\overline{\lambda} t}
$$
into Eq.~(\ref{Eq:Linear}) and equate separately the terms at $e^{\lambda t}$ and $e^{\overline{\lambda} t}$. Thus, we obtain a spectral problem
\begin{eqnarray}
\lambda v_+ &=& \mu v_+
+ \fr{\kappa_\mathrm{v}}{2} \mathcal{K}_\mathrm{v} ( v_+ - v_- )
+ \fr{i\kappa_\mathrm{s}}{2\pi} \mathcal{K}_\mathrm{s} ( v_+ + v_- ),
\label{EVP:FP:1}\\[2mm]
\lambda v_- &=& \overline{\mu} v_-
- \fr{\kappa_\mathrm{v}}{2} \mathcal{K}_\mathrm{v} ( v_+ - v_- )
- \fr{i\kappa_\mathrm{s}}{2\pi} \mathcal{K}_\mathrm{s} ( v_+ + v_- ).
\label{EVP:FP:2}
\end{eqnarray}
Now, we analyze the properties of spatial Fourier modes
$$
(v_+(x),v_-(x))^\mathrm{T} = (V_+,V_-)^\mathrm{T} e^{i m x}\quad\mbox{with}\quad m\in\mathbb{Z}\quad\mbox{and}\quad(V_+,V_-)^\mathrm{T}\in\mathbb{C}^2.
$$
Inserting this ansatz into Eqs.~(\ref{EVP:FP:1}), (\ref{EVP:FP:2}) and using the identities
$$
\mathcal{K}_\mathrm{v} e^{i m x} = 2\pi \hat{W}_{\mathrm{v},m} e^{i m x}
\quad\mbox{and}\quad
\mathcal{K}_\mathrm{s} e^{i m x} = 2\pi \hat{W}_{\mathrm{s},m} e^{i m x}
$$
with the coefficients $\hat{W}_{\mathrm{v},m}$ and $\hat{W}_{\mathrm{s},m}$
defined by~(\ref{Fourier:W}), we obtain a linear system
\begin{eqnarray*}
\lambda V_+ &=& \mu V_+
+ \pi \kappa_\mathrm{v} \hat{W}_{\mathrm{v},m} ( V_+ - V_- )
+ i\kappa_\mathrm{s} \hat{W}_{\mathrm{s},m} ( V_+ + V_- ),\\[2mm]
\lambda V_- &=& \overline{\mu} V_-
- \pi \kappa_\mathrm{v}  \hat{W}_{\mathrm{v},m} ( V_+ - V_- )
- i\kappa_\mathrm{s} \hat{W}_{\mathrm{s},m} ( V_+ + V_- ).
\end{eqnarray*}
Its characteristic equation reads
$$
\lambda^2 - 2 \lambda \Real( \mu + P_m ) + |\mu + P_m|^2 - |P_m|^2 = 0
$$
where
$$
P_m = \pi \kappa_\mathrm{v} \hat{W}_{\mathrm{v},m} + i\kappa_\mathrm{s} \hat{W}_{\mathrm{s},m}.
$$
Solving this equation, we obtain
\begin{eqnarray*}
\lambda_\pm &=& \Real( \mu + P_m ) \pm \sqrt{ |P_m|^2 - \left(  \Imag( \mu + P_m ) \right)^2 } \\[2mm]
&=& \Real( \mu ) + \pi \kappa_\mathrm{v} \hat{W}_{\mathrm{v},m}
\pm \sqrt{ \pi^2 \kappa_\mathrm{v}^2 \hat{W}_{\mathrm{v},m}^2 + \kappa_\mathrm{s}^2 \hat{W}_{\mathrm{s},m}^2 - \left(  \Imag( \mu ) + \kappa_\mathrm{s} \hat{W}_{\mathrm{s},m} \right)^2 }.
\end{eqnarray*}
Since the Fourier coefficients of $W_\mathrm{v}(x)$ and $W_\mathrm{s}(x)$
always satisfy the decay property
$|\hat{W}_{\mathrm{v},m}|,|\hat{W}_{\mathrm{s},m}|\to 0$ for $|m|\to\infty$,
this means $\lambda_\pm\to\mu$ or $\lambda_\pm\to\overline{\mu}$ for $|m|\to\infty$.
In other words, any potential instability of $u_0$ can be associated
with relatively small indices $m$ in the above formula for $\lambda_\pm$.

\begin{figure}[h]
\includegraphics[height=0.3\textwidth]{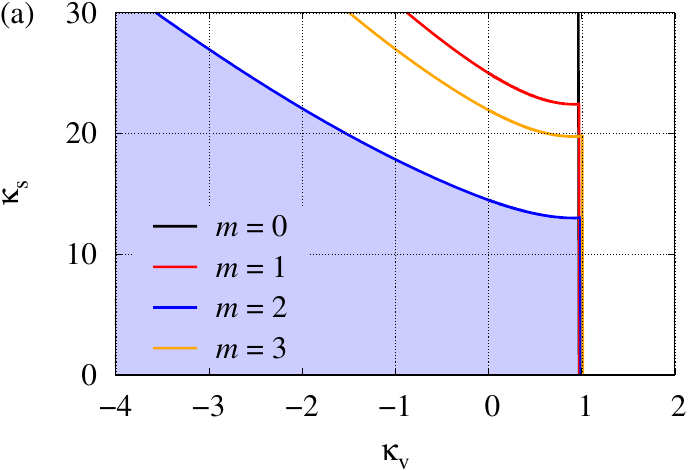}
\hspace{5mm}
\includegraphics[height=0.3\textwidth]{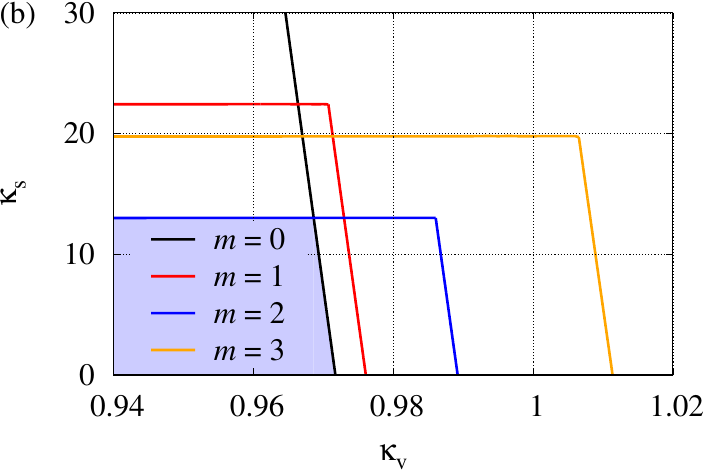}
\caption{
(a) Stability boundaries of the uniform state of Eq.~(\ref{Eq:MeanField})
determined by the condition $\max( \Real\lambda_+, \Real\lambda_- ) = 0$
for $m = 0,1,2$ and $3$. The uniform state is stable in the shaded region.
(b)~Enlargement of the range around $\kappa_\mathrm{v} = 1$.
Note that as $\kappa_s$ increases above $13.0$,
the pair of complex conjugate eigenvalues for $m=2$
turns into a pair of real eigenvalues,
so that the corresponding stability boundary has a corner point.
Other parameters: $\eta_0 = 1$ and $\gamma = 0.5$.}
\label{Fig:Stability}
\end{figure}

Instabilities of the uniform state for coupling functions~(\ref{Wv:Example}), (\ref{Ws:Example}) and $m=0,1,2,3$ are shown in Fig.~\ref{Fig:Stability}.
Their position in the $(\kappa_\mathrm{v},\kappa_\mathrm{s})$-plane
indicates that only the modes with $m = 0$ and $m = 2$
determine stability boundaries. Note that along the curve $m = 0$,
the critical eigenvalues everywhere have non-zero imaginary parts
(at least for the range $\kappa_\mathrm{s}\in[0,30]$).
In contrast, on the curve $m = 2$ we find this property
only for $\kappa_\mathrm{s} < 13.0$.
Above this value, the pair of complex conjugate eigenvalues
turns into a pair of real eigenvalues,
so that the corresponding stability boundary has a corner point.
As a result, two qualitatively different bifurcation scenarios of the uniform state
are expected for $\kappa_\mathrm{s} < 13.0$ and $\kappa_\mathrm{s} > 13.0$.
Indeed, for $\kappa_\mathrm{s}=10$ the uniform state with $\Real(u) \ge 0$
is stable to the left of the $m=0$ curve and as $\kappa_\mathrm{v}$ increases,
a Hopf bifurcation occurs, which leads to the creation
of a time-dependent (but still spatially uniform) solution.
However, for $\kappa_\mathrm{s}=20$ the uniform state loses stability to the $m=2$ mode as $\kappa_\mathrm{v}$ is increased.
The eigenvalues corresponding to this instability are real, so this is a Turing bifurcation
creating a stationary spatial pattern.

%%%%%%%%%
%%%%%%%%%%%%%%

\subsection{Stationary states}
\label{Sec:SS}

Stationary spatial patterns are non-constant time-independent solutions of Eq.~(\ref{Eq:MeanField}), see Fig.~\ref{Fig:Ks20_00}(a).
We can find them by solving a discretized version of Eq.~(\ref{Eq:MeanField}).
Alternatively, we can find them using a self-consistency equation we are about to derive.
For this, we insert the ansatz $u = a(x)$ into Eq.~(\ref{Eq:MeanField}) and obtain
\begin{equation}
\gamma + i F(x) - \kappa_\mathrm{v} a(x) - i a^2(x) = 0,
\label{Eq:FP:SS}
\end{equation}
where
\begin{equation}
F(x) = \eta_0 + \kappa_\mathrm{v} \mathcal{K}_\mathrm{v} \Imag(a) + \fr{\kappa_\mathrm{s}}{\pi} \mathcal{K}_\mathrm{s} \Real(a).
\label{Eq:F:SS}
\end{equation}
Using Proposition~\ref{Proposition:FP:General},
we can solve Eq.~(\ref{Eq:FP:SS}) for each $x$ separately
\begin{equation}
a(x) = \mathcal{S}( F(x), \kappa_\mathrm{v}, \gamma ).
\label{Eq:F_to_a}
\end{equation}
Then, inserting the result into Eq.~(\ref{Eq:F:SS}),
we obtain a self-consistency equation
\begin{equation}
F(x) = \eta_0
+ \kappa_\mathrm{v} \mathcal{K}_\mathrm{v} \Imag\left( \mathcal{S}\left( F(x), \kappa_\mathrm{v}, \gamma \vphantom{\sum} \right) \right)
+ \fr{\kappa_\mathrm{s}}{\pi} \mathcal{K}_\mathrm{s} \Real\left( \mathcal{S}\left( F(x), \kappa_\mathrm{v}, \gamma \vphantom{\sum} \right) \right).
\label{Eq:SC:SS}
\end{equation}
Note that Eq.~(\ref{Eq:SC:SS}) should be solved
with respect to the unknown function $F(x)$, which, in turn, determines
the corresponding stationary state $a(x)$ by formula~(\ref{Eq:F_to_a}).
By construction this procedure allows us to find all stationary states satisfying $|a(x)|\le 1$,
while every unphysical state that violates this inequality for at least some $x$ is automatically discarded.

In general, Eq.~(\ref{Eq:SC:SS}) cannot be solved exactly but only approximately.
Given that $F(x)$ is real and $2\pi$-periodic,
it is convenient to approximate it using a truncated Fourier series 
\begin{equation}
F(x) = \sum\limits_{m=0}^{2M} \hat{f}_m \phi_m(x)\quad\mbox{with}\quad\hat{f}_m\in\mathbb{R},
\label{Ansatz:F:Stationary}
\end{equation}
where
\begin{align}
\phi_0(x) & = 1, \nonumber \\
\phi_m(x) & = \sqrt{2} \cos( (m+1) x / 2 ) \:\: \mbox{for odd}\:\: m, \label{eq:basis} \\
\phi_m(x) & = \sqrt{2} \sin( m x / 2 )  \:\:\mbox{for even}\:\: m, \nonumber
\end{align}
are elements of the $L^2$-orthogonal trigonometric basis.
Inserting this ansatz into Eq.~(\ref{Eq:SC:SS}) and projecting the resulting relation
onto the basis function $\phi_m(x)$, we obtain a system of equations
\begin{equation}
\hat{f}_m = \eta_0 \delta_{m0} + \int_0^{2\pi} \left[ \kappa_\mathrm{v} \hat{W}_{\mathrm{v},m'} \Imag\left( \mathcal{S}\left( F(x), \kappa_\mathrm{v}, \gamma \vphantom{\sum} \right) \right)
+ \fr{\kappa_\mathrm{s}}{\pi} \hat{W}_{\mathrm{s},m'} \Real\left( \mathcal{S}\left( F(x), \kappa_\mathrm{v}, \gamma \vphantom{\sum} \right) \right) \right] \phi_m(x) dx
\label{Eq:System:f:SS}
\end{equation}
for $m=0,1,2,\dots 2M$,
where $\delta_{mn}$ is the Kronecker delta,
$\hat{W}_{\mathrm{v},m}$ and $\hat{W}_{\mathrm{s},m}$
are the Fourier coefficients of $W_\mathrm{v}(x)$ and $W_\mathrm{s}(x)$
given by~(\ref{Fourier:W}), and
$$
m' = 0\:\:\mbox{if}\:\: m=0,\quad
m' = (m+1)/2\:\:\mbox{if}\:\: m\:\:\mbox{is odd},\quad
m' = m/2\:\:\mbox{if}\:\: m\:\:\mbox{is even}.
$$

For completeness, we list here four identities
\begin{eqnarray}
&&
\int_0^{2\pi} (\mathcal{K}_\mathrm{v} u)(x) \cos(m x) dx = 2\pi \hat{W}_{\mathrm{v},m}\int_0^{2\pi} u(x) \cos(m x) dx,
\label{Eq:Kv:cos:u}\\[2mm]
&&
\int_0^{2\pi} (\mathcal{K}_\mathrm{v} u)(x) \sin(m x) dx = 2\pi \hat{W}_{\mathrm{v},m}\int_0^{2\pi} u(x) \sin(m x) dx,
\label{Eq:Kv:sin:u}\\[2mm]
&&
\int_0^{2\pi} (\mathcal{K}_\mathrm{s} u)(x) \cos(m x) dx = 2\pi \hat{W}_{\mathrm{s},m}\int_0^{2\pi} u(x) \cos(m x) dx,
\label{Eq:Ks:cos:u}\\[2mm]
&&
\int_0^{2\pi} (\mathcal{K}_\mathrm{s} u)(x) \sin(m x) dx = 2\pi \hat{W}_{\mathrm{s},m}\int_0^{2\pi} u(x) \sin(m x) dx,
\label{Eq:Ks:sin:u}
\end{eqnarray}
which are valid for arbitrary real function $u(x)$, any integer $m$,
and any integral operators~$\mathcal{K}_\mathrm{v}$ and~$\mathcal{K}_\mathrm{s}$
with symmetric (even) kernels $W_\mathrm{v}(x)$ and $W_\mathrm{s}(x)$, respectively.
We used these identities to obtain formula~(\ref{Eq:System:f:SS}).

Note that for symmetric kernels $W_\mathrm{v}(x)$ and $W_\mathrm{s}(x)$
the dimensionality of system~(\ref{Eq:System:f:SS}) can be halved.
Indeed, it is easy to see that in this case
the operators~$\mathcal{K}_\mathrm{v}$ and~$\mathcal{K}_\mathrm{s}$
are invariant on the subspace of even functions.
Therefore, we can assume that $F(x)$ is even
and hence $\hat{f}_m = 0$ for all even positive $m$.
Importantly, after such an assumption
we do not need to care about the translational symmetry of Eq.~(\ref{Eq:SC:SS}),
which would otherwise require the addition of a pinning condition to~(\ref{Eq:System:f:SS}).

In conclusion, we make a remark that for each stationary pattern $a(x)$
found using the self-consistency equation~(\ref{Eq:SC:SS}),
its linear stability can be analyzed
according to the scheme of Section~\ref{Sec:US:LS}.
For brevity, we do not perform such an analysis here,
but note that conceptually this will not be very different
from the linear stability analysis of bump states described in~\cite{omelai22}.

\subsection{Travelling wave}
\label{sec:TW}

In this section, we consider the traveling wave solutions of Eq.~(\ref{Eq:MeanField}),
which are uniformly drifting spatial patterns, see Fig.~\ref{Fig:TW}(a),
with a mathematical form
$$
u = a(x + s t)\quad\mbox{where}\quad s\ne 0.
$$

Our primary approach to studying them is to realise that such waves
are stationary
in a uniformly travelling coordinate system, travelling at the speed of the wave.
Thus they are steady states of
\be
 \pf{a}{t} = \gamma - \kappa_\mathrm{v} a + i \left[ \eta_0 + \kappa_\mathrm{v} \mathcal{K}_\mathrm{v} \Imag(a) + \fr{\kappa_\mathrm{s}}{\pi} \mathcal{K}_\mathrm{s} \Real(a) - a^2 \right]-s \pf{a}{x} \label{eq:trav}
\ee
where $s$ is the speed at which they travel~\cite{laiome20,lai14b}. Thus we can find them by
uniformly discretising~\eqref{eq:trav} in space and solving the corresponding large set of
coupled algebraic equations using Newton's method. The spatial derivative can be approximated using
finite differences. One advantage of this method is that the stability of the wave
can be determined from the eigenvalues of the linearisation of~\eqref{eq:trav} about
a steady state.

An alternative approach to the study of travelling waves in Eq.~(\ref{Eq:MeanField})
consists in deriving a self-consistency equation for them.
For this, we note that if $s > 0$ (which will be the default case below)
and $a(x)$ is a time-independent solution of Eq.~(\ref{eq:trav}),
then $a(x)$ satisfies
\begin{equation}
\df{a}{x} = \fr{\gamma}{s} + \fr{i F(x)}{s} - \fr{\kappa_\mathrm{v}}{s} a - \fr{i}{s} a^2,
\label{Eq:MeanField:TW}
\end{equation}
where
$$
F(x) = \eta_0 + \kappa_\mathrm{v} \mathcal{K}_\mathrm{v} \Imag(a) + \fr{\kappa_\mathrm{s}}{\pi} \mathcal{K}_\mathrm{s} \Real(a).
%\label{Def:F:TW}
$$
For Eq.~(\ref{Eq:MeanField:TW}) we know
(see Proposition~\ref{Proposition:Sln:w} in Appendix)
that for any $\gamma > 0$, $s > 0$,
any $\kappa_\mathrm{v}\in\mathbb{R}$ and any real $2\pi$-periodic function $F(x)$,
this equation has one and only one $2\pi$-periodic solution,
which lies entirely in the right half-plane $\mathbb{P} = \{ w\in\mathbb{C}\::\: \Real\:w> 0 \}$.
Therefore, we can write
\begin{equation}
a(x) = \mathcal{U}(F(x), \kappa_\mathrm{v}, \gamma, s),
\label{Eq:F_to_a:TW}
\end{equation}
where $\mathcal{U}(\cdot)$ denotes the corresponding solution operator.
To be consistent with the above definition of $F(x)$, we must have
\begin{equation}
F(x) = \eta_0
+ \kappa_\mathrm{v} \mathcal{K}_\mathrm{v} \Imag\left( \mathcal{U}\left( F(x), \kappa_\mathrm{v}, \gamma, s \vphantom{\sum} \right) \right)
+ \fr{\kappa_\mathrm{s}}{\pi} \mathcal{K}_\mathrm{s} \Real\left( \mathcal{U}\left( F(x), \kappa_\mathrm{v}, \gamma, s \vphantom{\sum} \right) \right).
\label{Eq:SC:TW}
\end{equation}
Thus, in order to find a traveling wave solution of Eq.~(\ref{Eq:MeanField}),
we can first solve Eq.~(\ref{Eq:SC:TW}) for $F(x)$
and then use~(\ref{Eq:F_to_a:TW}) to calculate the corresponding wave profile.
Note that due to the translational symmetry of Eq.~(\ref{Eq:MeanField})
we need to add an additional pinning condition such as
\begin{equation}
\int_0^{2\pi} F(x) \sin x\: dx = 0
\label{Eq:Pinning:TWa}
\end{equation}
to pick up a unique solution of this equation.
This condition also allows us to find the speed of traveling wave $s$.

Let us make a few remarks about the numerical implementation
of the method based on the self-consistency equation~(\ref{Eq:SC:TW}).
Recall that the algorithm for calculating the solution operator $\mathcal{U}(\cdot)$
is described in Remark~\ref{remark2} in Appendix.
Moreover, the oprtimal strategy for solving Eq.~(\ref{Eq:SC:TW})
is the application of the Galerkin method.
For this, we insert the ansatz~(\ref{Ansatz:F:Stationary}) into Eq.~(\ref{Eq:SC:TW})
and project the resulting relation onto the trigonometric basis $\phi_m(x)$ 
defined in~\eqref{eq:basis}.
As a result we obtain a system of $2 M + 1$ equations
\begin{equation}
\hat{f}_m = \eta_0 \delta_{m0} + \int_0^{2\pi} \left[ \kappa_\mathrm{v} \hat{W}_{\mathrm{v},m'} \Imag\left( \mathcal{U}\left( F(x), \kappa_\mathrm{v}, \gamma, s \vphantom{\sum} \right) \right)
+ \fr{\kappa_\mathrm{s}}{\pi} \hat{W}_{\mathrm{s},m'} \Real\left( \mathcal{U}\left( F(x), \kappa_\mathrm{v}, \gamma, s \vphantom{\sum} \right) \right) \right] \phi_m(x) dx
\label{Eq:System:f:TW}
\end{equation}
where
$$
m' = 0\:\:\mbox{if}\:\: m=0,\quad
m' = (m+1)/2\:\:\mbox{if}\:\: m\:\:\mbox{is odd},\quad
m' = m/2\:\:\mbox{if}\:\: m\:\:\mbox{is even}.
$$
Note that the solutions of Eq.~(\ref{Eq:SC:TW}) do not necessarily
have to be reflection-symmetric, so all integer indices $m=0,1,\dots,2M$
must be taken into account in system~(\ref{Eq:System:f:TW}).
Moreover, in this case the pinning condition~(\ref{Eq:Pinning:TWa}) cannot be omitted,
although it becomes particularly simple
\begin{equation}
\hat{f}_2 = 0.
\label{Eq:Pinning:TW}
\end{equation}
Combining~(\ref{Eq:System:f:TW}) with~(\ref{Eq:Pinning:TW}),
we obtain a system that determines all $\hat{f}_m$ in~(\ref{Ansatz:F:Stationary})
and the wave speed $s$.

\subsection{Standing wave}
\label{sec:SW}

We use the term standing wave to refer to a spatiotemporal pattern
that is non-constant in the spatial direction
and exhibits periodic oscillation with the same collective period
for each position~$x$. Examples of such patterns
are shown in Fig.~\ref{Fig:SW:TW}(a), Fig.~\ref{Fig:SW}
and Fig.~\ref{Fig:Ks20_00}(b),(c).
Mathematically, each standing wave 
corresponds to spatially heterogeneous periodic solution of Eq.~(\ref{Eq:MeanField}),
i.e. to a solution of the form $u = a(x,t)$
which is $T$-periodic in $t$ for some $T > 0$.
To study how these periodic solutions depend on system parameters,
we can again use two approaches.
On the one hand, we can look for periodic solutions
of the discretized version of Eq.~(\ref{Eq:MeanField}).
On the other hand, we can derive a self-consistency equation
similar to Eq.~(\ref{Eq:SC:SS}) and Eq.~(\ref{Eq:SC:TW}).
To write such a self-consistency equation, we first perform time-rescaling,
by defining a new unknown function $v(x,t) = u(x,t/\omega)$
where $\omega = 2\pi / T$ is the cyclic frequency corresponding to period $T$.
The new function $v(x,t)$ satisfies
$$
\omega \pf{v}{t} = \gamma - \kappa_\mathrm{v} v + i \left[ \eta_0 + \kappa_\mathrm{v} \mathcal{K}_\mathrm{v} \Imag(v) + \fr{\kappa_\mathrm{s}}{\pi} \mathcal{K}_\mathrm{s} \Real(v) - v^2 \right],
$$
or equivalently
\begin{equation}
\pf{v}{t} = \fr{\gamma}{\omega} + \fr{i F(x,t)}{\omega} - \fr{\kappa_\mathrm{v}}{\omega} v - \fr{i}{\omega} v^2,
\label{Eq:MeanField_}
\end{equation}
where
\begin{equation}
F(x,t) =  \eta_0 + \kappa_\mathrm{v} \mathcal{K}_\mathrm{v} \Imag(v) + \fr{ \kappa_\mathrm{s} }{\pi} \mathcal{K}_\mathrm{s} \Real(v).
\label{Def:F:SW}
\end{equation}
For every fixed $x\in[0,2\pi]$, Eq.~(\ref{Eq:MeanField_}) has the same form
as Eq.~(\ref{Eq:MeanField:TW}), but with other variables
$x\mapsto t$ and $s\mapsto \omega$.
Therefore, using the notation of the solution operator $\mathcal{U}(\cdot)$,
we can write
\begin{equation}
v(x,t) = \mathcal{U}\left( F(x,t), \kappa_\mathrm{v}, \gamma, \omega \right).
\label{Eq:F_to_v}
\end{equation}
Inserting this into the definition of $F(x,t)$, we obtain a self-consistency equation
\begin{equation}
F(x,t) = \eta_0
+ \kappa_\mathrm{v} \mathcal{K}_\mathrm{v} \Imag\left( \mathcal{U}\left( F(x,t), \kappa_\mathrm{v}, \gamma, \omega \vphantom{\sum} \right) \right)
+ \fr{\kappa_\mathrm{s}}{\pi} \mathcal{K}_\mathrm{s} \Real\left( \mathcal{U}\left( F(x,t), \kappa_\mathrm{v}, \gamma, \omega \vphantom{\sum} \right) \right).
\label{Eq:SC}
\end{equation}
Recalling that every solution $v_0(x,t)$ of the periodic boundary value problem for Eq.~(\ref{Eq:MeanField_})
determines a two-parameter familily of solutions $v_0(x+x_0,t+t_0)$ with $x_0,t_0\in(0,2\pi)$,
we equip Eq.~(\ref{Eq:SC}) with two additional pinning conditions
\begin{equation}
\int_0^{2\pi} \int_0^{2\pi} F(x,t) \sin x\: dx\: dt = \int_0^{2\pi} \int_0^{2\pi} F(x,t) \sin t\: dx\: dt = 0,
\label{Eq:Pinning}
\end{equation}
which allow us to pick up a single function from this family.

In the following we use a Fourier sum approximation for $F(x,t)$:
\begin{equation}
F(x,t) = \sum\limits_{m=0}^{2M} \sum\limits_{n=0}^{2N} \hat{f}_{mn} \phi_m(x)\phi_n(t),
\label{Ansatz:F}
\end{equation}
where $\hat{f}_{mn}\in\mathbb{R}$, and the $\phi_m$ are given in~\eqref{eq:basis}.
%$$
%\phi_0(x) = 1,\quad
%\phi_m(x) = \sqrt{2} \cos( (m+1) x / 2 )\:\:\mbox{for odd}\:\: m,\quad
%\phi_m(x) = \sqrt{2} \sin( m x / 2 )\:\:\mbox{for even}\:\: m.
%$$
Applying the standard Galerkin scheme to Eq.~(\ref{Eq:SC}) we find
a system of algebraic equations that determines the coefficients $\hat{f}_{mn}$:
\begin{eqnarray}
\hat{f}_{mn} = \eta_0 \delta_{m0} \delta_{n0} &+& \left\langle \kappa_\mathrm{v} \mathcal{K}_\mathrm{v} \Imag\left( \mathcal{U}\left( F(x,t), \kappa_\mathrm{v}, \gamma, \omega \vphantom{\sum} \right) \right) \right.\nonumber\\[2mm]
&+& \left. \fr{\kappa_\mathrm{s}}{\pi} \mathcal{K}_\mathrm{s} \Real\left( \mathcal{U}\left( F(x,t), \kappa_\mathrm{v}, \gamma, \omega \vphantom{\sum} \right) \right), \phi_m(x)\phi_n(t) \right\rangle
\label{Eq:System:f}
\end{eqnarray}
where
$$
\langle u, v \rangle = \fr{1}{4\pi^2} \int_0^{2\pi} \int_0^{2\pi} u(x,t) \overline{v}(x,t) dx\:dt.
$$
Moreover, using the identities~(\ref{Eq:Kv:cos:u})--(\ref{Eq:Ks:sin:u}), we simplify Eq.~(\ref{Eq:System:f}) to obtain
\begin{eqnarray}
\hat{f}_{mn} = \eta_0 \delta_{m0} \delta_{n0} &+& \left\langle 2 \pi \kappa_\mathrm{v} \hat{W}_{\mathrm{v},m'} \Imag\left( \mathcal{U}\left( F(x,t), \kappa_\mathrm{v}, \gamma, \omega \vphantom{\sum} \right) \right) \right.\nonumber\\[2mm]
&+& \left. 2 \kappa_\mathrm{s} \hat{W}_{\mathrm{s},m'} \Real\left( \mathcal{U}\left( F(x,t), \kappa_\mathrm{v}, \gamma, \omega \vphantom{\sum} \right) \right), \phi_m(x)\phi_n(t) \right\rangle,
\label{Eq:System:f_}
\end{eqnarray}
where
$$
m' = 0\:\:\mbox{if}\:\: m=0,\quad
m' = (m+1)/2\:\:\mbox{if}\:\: m\:\:\mbox{is odd},\quad
m' = m/2\:\:\mbox{if}\:\: m\:\:\mbox{is even}.
$$

For the patterns shown in Fig.~\ref{Fig:SW:TW}(a) and Fig.~\ref{Fig:SW},
we can, without loss of generality, assume $u(-x,t) = u(x,t)$ and hence $F(-x,t) = F(x,t)$.
In this case, $\hat{f}_{mn} = 0$ for all even positive $m$.
Therefore, the first pinning condition in~(\ref{Eq:Pinning}) is satisfied automatically.
For the second pinning condition, inserting the ansatz~(\ref{Ansatz:F}) into it,
we obtain
\begin{equation}
\hat{f}_{02} = 0.
\label{Eq:Pinning_}
\end{equation}
In summary, solving the system~(\ref{Eq:System:f_}), (\ref{Eq:Pinning_})
we are able to find an approximate solution $F(x,t)$ of the form~(\ref{Ansatz:F})
as well as the corresponding frequency $\omega$.
Then, by formula~(\ref{Eq:F_to_v}) we can calculate the solution of Eq.~(\ref{Eq:MeanField_}),
which after the appropriate time rescaling $u(x,t) = v(x,\omega t)$ yields the standing wave
solution of Eq.~(\ref{Eq:MeanField}).

\subsection{Lurching solutions}

In this section, we consider lurching waves ---
the most complex type of solutions of Eq.~(\ref{Eq:MeanField})
that we have been able to investigate analytically.
A lurching wave is a travelling wave,
but rather than having a constant profile, its profile periodically oscillates;
see Fig.~\ref{Fig:SW:TW}(b), Fig.~\ref{Fig:TW}(b) and Fig.~\ref{Fig:Ks20_00}(d).
These are sometimes referred to as modulated travelling waves~\cite{sansch01}.
Each such wave can be characterized as
a fixed point of a ``shift and run'' map of the type described in~\cite{wascis10}, but
with a continuous shift rather than the discrete one used in~\cite{wascis10}.
More precisely, we look for lurching waves as fixed points of the map
\[
   u_{n+1}=P(u_n;\chi,T)
\]
where $P(u;\chi,T)$ is defined by taking the function $u(x)$, shifting it in space by an amount
$\chi$ (to the left if the pattern is moving to the right,
recalling that the domain is periodic) and then integrating~\eqref{Eq:MeanField} with
this shifted pattern as the initial condition for an amount of time $T$. A continuous
shift is easily implemented in Fourier space. If
\[
   u(x)=\sum_{n=-M}^M a_ne^{inx}
\]
then
\[
   u(x+\chi)=\sum_{n=-M}^M \hat{a}_ne^{inx}
\]
where $\hat{a}_n=a_ne^{in\chi}$. Note that the stability of a lurching wave can be found
from the eigenvalues of the Jacobian of the map $P$ evaluated at the relevant fixed 
point~\cite{wascis10}.

It is natural to expect that lurching waves can also be described
using a variant of the self-consistency approach.
However, so far we have not been able to do this.
The main problem is that after time rescaling $v(x,t) = u(x,t/\omega)$
we need to consider Eq.~(\ref{Eq:MeanField_}) on the functional space
$$
\tilde{C}_\chi = \left\{ v\in C([0,2\pi]^2; \mathbb{C})\::\:
v(x + 2\pi, t) = v(x, t),\quad
v(x + \chi, t + 2\pi) = v(x,t) \right\}
$$
and show that for every $F\in \tilde{C}_\chi$ equation~(\ref{Eq:MeanField_})
has a unique solution $v\in \tilde{C}_\chi$ satisfying $|v|\le 1$.
This turns out to be a nontrivial task due to the nonlocal character
of the boundary condition $v(x + \chi, t + 2\pi) = v(x,t)$.
So we leave it for future investigation.

\section{Results}
\label{sec:results}

Now we show how the analytical methods developed in Section~\ref{sec:bif}
can be used to explain the empirical stability diagrams from Section~\ref{sec:phen}.
We begin by reviewing the computational efficiency
of the two approaches described in Section~\ref{sec:bif}.
Then, we show the calculated bifurcation diagrams associated with Figures~\ref{Fig:Scan}
and~\ref{Fig:Scan:Ks20_00}, which make visible the relationships between
various spatiotemporal patterns which are solutions of Eq.~(\ref{Eq:MeanField}),
sometimes via unstable (and therefore invisible) solutions of this equation.

\subsection{Computational efficiency of the proposed methods}
\begin{itemize}
\item {\bf Stationary states.} To find stationary states of~\eqref{Eq:MeanField} one could
uniformly discretise this in space and solve the corresponding large set of
coupled algebraic equations using Newton's method. Or one could solve the 
system~\eqref{Eq:System:f:SS}. Which method is more efficient depends on the number of spatial
points used (call this $\hat{N}$) and $M$, the number of spatial harmonics used to approximate
$F(x)$. If $\hat{N}$ is small and $M$ is large the former method is more efficient, but
if $\hat{N}$ is large and $M$ is small the latter method is more efficient. The former method
has the advantage that the stability can be readily found from the eigenvalues of the linearisation
about the stationary state.

\item {\bf Travelling waves.} It was found to be much more efficient to use the approach discussed
at the beginning of Sec.~\ref{sec:TW}, i.e.~equation~\eqref{eq:trav}, than to solve the system
of equations~\eqref{Eq:System:f:TW}. The main reasons for this are (i) the large number of
harmonics used to approximate $F$: we used $M=50$, giving 101 unknowns, and (ii) the fact that 
to evaluate the operator $\mathcal{U}$ once we need to numerically integrate~\eqref{Eq:MeanField:TW} 
four times, as explained in Remark~\ref{remark2}. Note that if the kernels $W_\mathrm{v}$ and 
$W_\mathrm{s}$ were described exactly by a small number of sinusoidal functions, $F$ could be
expressed exactly as a finite Fourier series, with a corresponding small number of
unknowns~\cite{ome23,LaiO2023}.

\item {\bf Standing waves.} 
A standing wave is a periodic solution of~\eqref{Eq:MeanField} 
and so could be studied using conventional methods after spatial discretisation~\cite{lai14b}.
This method was found to be much more efficient than the method proposed in Sec.~\ref{sec:SW},
due to the large number of unknowns required for that method, $(2M+1)(2N+1)$, and the need
to integrate~\eqref{Eq:MeanField_} four times, as explained above. 
\end{itemize}

\subsection{Bifurcation diagrams for $\kappa_\mathrm{s} = 10$}

\begin{figure}[t]
\includegraphics[width=0.75\textwidth]{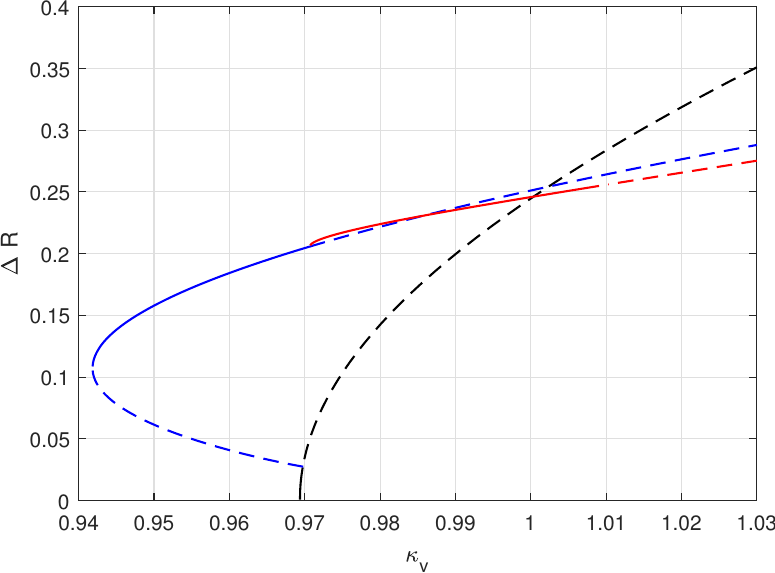}
\caption{Bifurcation diagram of standing waves for $\eta_0 = 1$, $\gamma = 0.5$
and $\kappa_\mathrm{s} = 10$. Black: spatially-uniform states. Blue: standing waves.
Red: standing waves with less symmetry. Solid: stable; dashed: unstable.}
\label{Fig:Ks10_00:SW:Theory}
\end{figure}

Recall that we found two coexisting stable solutions at $\kappa_\mathrm{v}=1$.
We first discuss the standing waves
and other related solutions of Eq.~(\ref{Eq:MeanField}).
The spatially-uniform state is stable for $\kappa_\mathrm{v} < 0.96934$ 
and it undergoes a supercritical Hopf bifurcation with $m=0$ at this point.
The emerging uniformly oscillating state is shown in black in
Fig.~\ref{Fig:Ks10_00:SW:Theory} and is stable upon creation but 
undergoes a subcritical bifurcation at $\kappa_\mathrm{v} = 0.96974$, 
generating an unstable branch of standing waves, shown in blue in Fig.~\ref{Fig:Ks10_00:SW:Theory}.
(This branch of spatially uniform oscillating states is stable only over a very small range of
$\kappa_\mathrm{v}$ values and was not observed in Fig.~\ref{Fig:Scan}.)
At $\kappa_\mathrm{v} = 0.94185$ a saddle-node bifurcation occurs
creating stable and unstable branches of standing waves.
At $\kappa_\mathrm{v} = 0.97069$ the branch of standing waves
undergoes a symmetry breaking bifurcation
and a branch of stable standing waves with more complex behavior and less symmetry 
(shown in red Fig.~\ref{Fig:Ks10_00:SW:Theory}) appears.
The latter loses its stability at $\kappa_\mathrm{v} = 1.0091$.
This analysis provide an explanation for the results shown in the left column of Fig.~\ref{Fig:Scan}:
there is a stable standing wave for smaller values of $\kappa_\mathrm{v}$
and one with less symmetry for larger values.

%%%%%%%%%%%%%%%%%%%
%%%%%%%%%%%%%%%

\begin{figure}[h]
\includegraphics[width=0.75\textwidth]{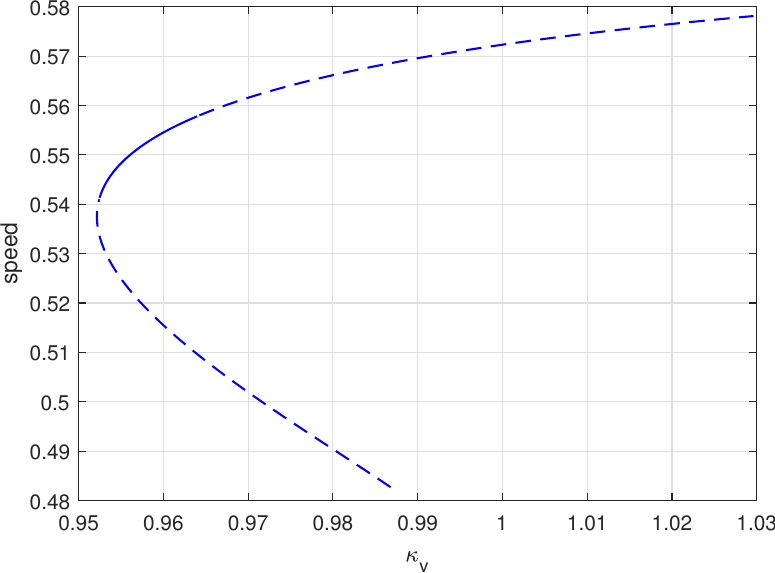}
\caption{Bifurcation diagram of travelling waves for $\eta_0 = 1$, $\gamma = 0.5$
and $\kappa_\mathrm{s} = 10$. Solid: stable; dashed: unstable. The two bifurcations at which the
wave loses stability are Hopf bifurcations.}
\label{Fig:Ks10_00:TW:Theory}
\end{figure}

We now discuss the travelling waves.
The travelling wave is stable at $\kappa_\mathrm{v} = 0.96$
and either decreasing or increasing $\kappa_\mathrm{v}$ destabilises the wave in a Hopf
bifurcation; see Fig.~\ref{Fig:Ks10_00:TW:Theory}. 
The leftmost bifurcation, at $\kappa_\mathrm{v} = 0.95243$, seems subcritical
and occurs to the right of the saddle-node bifurcation. Following the branch through the saddle-node
bifurcation it terminates in a collision with the spatially unform state at 
$\kappa_\mathrm{v} = 0.9868$. Put another way, the spatially uniform state undergoes 
a subcritical Hopf bifurcation with $m = 2$ at this parameter value, leading to the creation
of an unstable travelling wave --- see Fig.~\ref{Fig:Stability}.

%%%%%%%%%%%%%%%%%%
%%%%%%%%%%%%%%%%%%%%

\begin{figure}[h]
\includegraphics[width=0.85\textwidth]{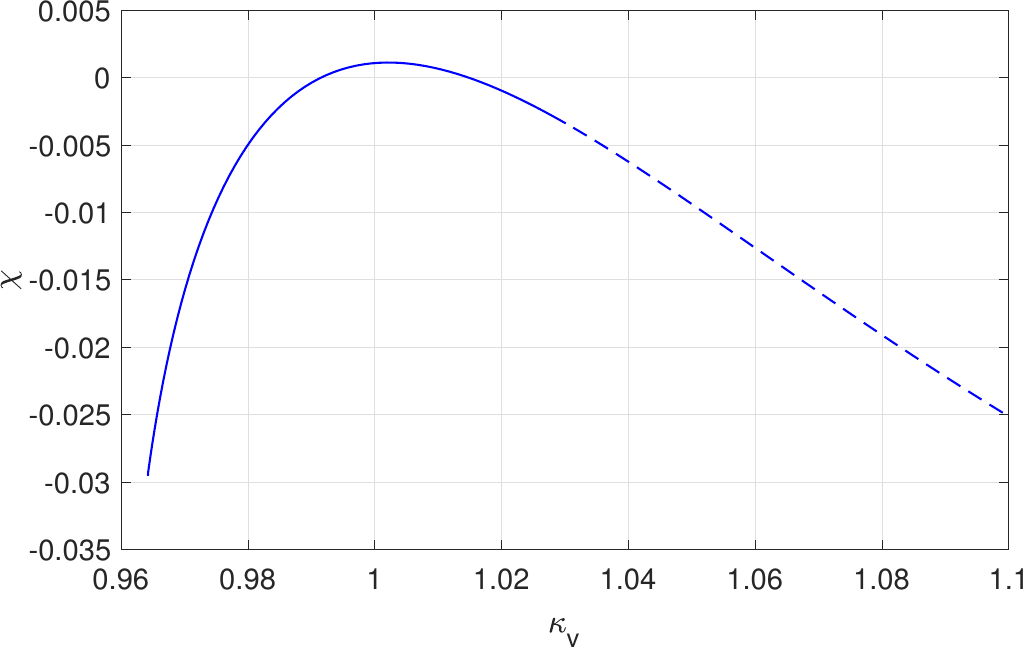}
\caption{Behaviour of lurching waves for $\eta_0 = 1$, $\gamma = 0.5$
and $\kappa_\mathrm{s} = 10$. $\chi$ is the spatial shift. Solid: stable; dashed: unstable.
The left end of the branch of solutions is created in the rightmost bifurcation shown
in Fig.~\ref{Fig:Ks10_00:TW:Theory}.}
\label{Fig:Ks10_00:TW:TheoryA}
\end{figure}

The rightmost Hopf bifurcation of the travelling wave, at $\kappa_\mathrm{v} = 0.96398$,
is supercritical and results in the creation of a stable lurching solution. 
We followed this lurching wave and a plot of the spatial shift
$\chi$ versus $\kappa_\mathrm{v}$ is shown in
Fig.~\ref{Fig:Ks10_00:TW:TheoryA}. The magnitude of $\chi$ is small, in keeping with
the slow drift mentioned in Sec.~\ref{sec:phen10}.
These lurching waves remain stable until $\kappa_\mathrm{v} = 1.0301$,
where they apparently undergo a torus bifurcation, creating a solution which is quasiperiodic
in a uniformly travelling coordinate frame.
As $\kappa_\mathrm{v}$ grows further, these waves become chaotic.
This analysis provide an explanation for the results shown in the right column of Fig.~\ref{Fig:Scan}:
the travelling wave in Fig.~\ref{Fig:Ks10_00:TW:Theory} has $\Delta R=0$, the lurching wave is seen 
at $\kappa_\mathrm{v} = 1$, for example, and the quasiperiodic and chaotic behaviour is seen in
Fig.~\ref{Fig:Scan}(d).

\subsection{Bifurcation diagram for $\kappa_\mathrm{s} = 20$.}

\begin{figure}[h]
\includegraphics[width=0.8\textwidth]{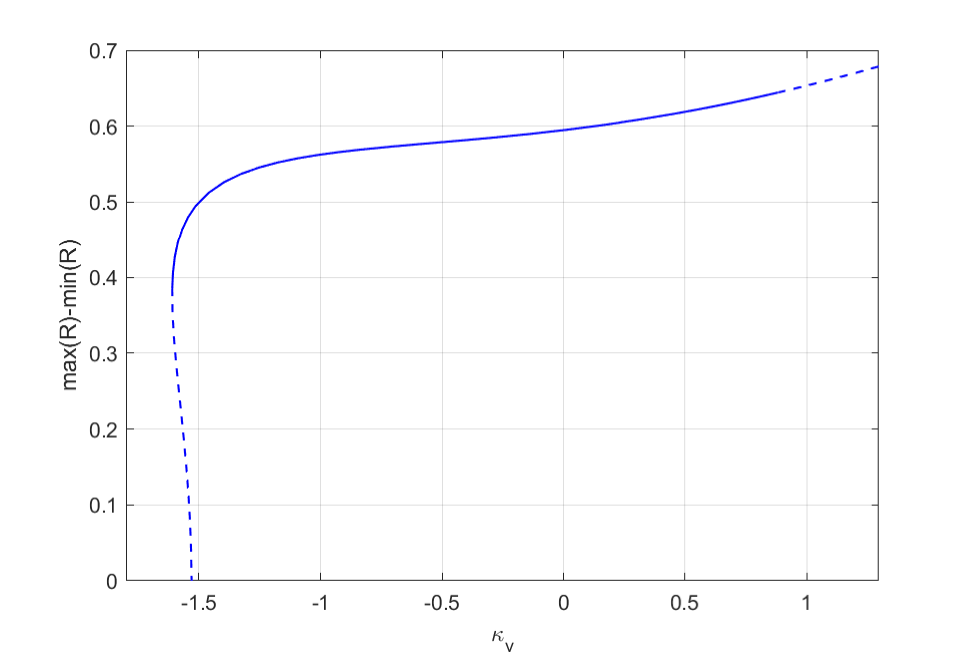}
\caption{Stationary two-bump solution $\eta_0 = 1$, $\gamma = 0.5$
and $\kappa_\mathrm{s} = 20$. Solid: stable; dashed: unstable.
The vertical axis shows $\max_x R(x)-\min_x R(x)$. The solution is stable between the
saddle-node bifurcation and a supercritical Hopf bifurcation at $\kappa_\mathrm{v} = 0.88565$.
(Note that negative values of $\kappa_\mathrm{v}$ are unphysical, but we plot them to show the
full branch of solutions.)}
\label{Fig:stat}
\end{figure}

\begin{figure}[h]
\includegraphics[width=0.8\textwidth]{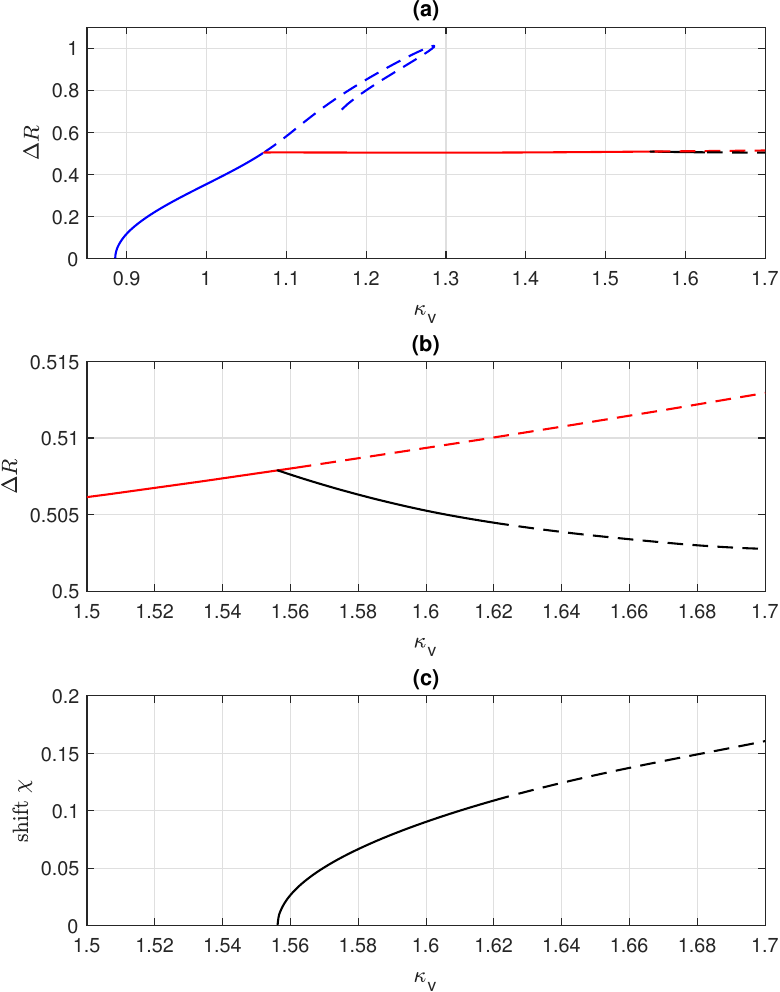}
\caption{Bifurcation diagrams for $\kappa_\mathrm{s} = 20$.
Blue: periodic solution. Red:
period-doubled solution. Black: lurching solution.
 Solid: stable; dashed: unstable. Panel (b) is an enlargment of panel (a). Panel (c)
shows the shift, $\chi$, of the lurching solution.}  
\label{Fig:Ks20_00:Theory}
\end{figure}

The uniform state undergoes a subcritical Turing bifurcation with $m=2$ 
at $\kappa_\mathrm{v} = -1.53$ (see Fig.~\ref{Fig:Stability}). An unstable branch
of stationary two-bump solutions emerges at this point, shown dashed in Fig.~\ref{Fig:stat}.
This solution becomes stable in a saddle-node bifurcation at $\kappa_\mathrm{v} = -1.6099$.
At $\kappa_\mathrm{v} = 0.88565$ a supercritical Hopf bifurcation occurs
and a branch of standing waves (or breathing two-bump solutions), shown in blue in
Fig.~\ref{Fig:Ks20_00:Theory} emerges.

These standing waves are stable until a symmetry breaking period-doubling bifurcation
at $\kappa_\mathrm{v} = 1.0719$. To the right from this point,
a branch of standing waves in the form of alternating two-bump patterns is stable, shown in solid red
in Fig.~\ref{Fig:Ks20_00:Theory}.
At $\kappa_\mathrm{v} = 1.5563$ this alternating pattern becomes unstable
through a supercritical torus bifurcation and a branch of lurching states appears,
shown in black in Fig.~\ref{Fig:Ks20_00:Theory}.
The lurching state is stable until $\kappa_\mathrm{v} = 1.6248$.
After that it undergoes a sequence of bifurcations leading to a chaotic regime.
Note that the branch of periodic solutions shown in blue in Fig.~\ref{Fig:Ks20_00:Theory}
terminates in a collision with an unstable spatially uniform periodic solution (not shown
or analysed). This analysis provides an explanation for the results shown 
in Fig.~\ref{Fig:Scan:Ks20_00}.

\section{Discussion}
\label{sec:disc}

In this paper, we considered the neural field equation~(\ref{Eq:MeanField}),
which describes the long-term dynamics of a ring network of QIF neurons
with nonlocal synaptic and gap junction coupling, in the limit of a large number of neurons.
We showed that apart from spatially uniform states,
this model also supports various spatiotemporal patterns,
including stationary bump states, standing and travelling waves,
and lurching waves. We showed how each of these states
can be analyzed semi-analytically using the discretized version of Eq.~(\ref{Eq:MeanField})
or using the corresponding self-consistency equation. We believe this to be the first
study of spatially extended QIF networks using self-consistency arguments.
Finally, we computed detailed bifurcation diagrams for~\eqref{Eq:MeanField}, which linked
the previous observations together.

The linear stability analysis of constant solutions of Eq.~(\ref{Eq:MeanField})
performed in Sec.~\ref{Sec:Uniform} revealed
that there are two qualitatively different bifurcation scenarios
for small and large values of synaptic coupling; see Fig.~\ref{Fig:Stability}.
More specifically, for $\kappa_\mathrm{s} < 13$
the uniform state loses its stability due to the Hopf bifurcation
with a spatially uniform eigenfunction, resulting in the creation
of a spatially-uniform oscillating state.
In contrast, for $\kappa_\mathrm{s} > 13$ the destabilization
of the uniform state occurs due to a Turing bifurcation
with a spatially modulated eigenfunction with wave number $m=2$.
This results in the creation of a ``two-bump'' stationary solution.

Using the algorithms described in Sec.~\ref{sec:bif},
we followed the five types of solutions of Eq.~(\ref{Eq:MeanField})
primarily as the parameter $\kappa_\mathrm{v}$ was varied.
This allowed us to better understand the relationship between these solutions
and provide an explanation of the numerical results found in Sec.~\ref{sec:phen}.
More specifically,
for $\kappa_\mathrm{s}=10$ we found coexisting standing waves and travelling waves,
and a type of lurching wave that resulted from a travelling wave undergoing an oscillatory
instability. For $\kappa_\mathrm{s}=20$ we found the scenario shown in Fig.~\ref{Fig:Ks20_00:Theory},
where more complex solutions are created as $\kappa_\mathrm{v}$ is increased. Interestingly,
the lurching solution found for this value of $\kappa_\mathrm{s}$ results from a standing wave
solution losing stability and starting to travel, hence the increase in $\chi$ from zero
as $\kappa_\mathrm{v}$ is increased, shown in Fig.~\ref{Fig:Ks20_00:Theory}(c).

The model~\eqref{Eq:MeanField} has many parameters and we have only investigated varying one
of them for several values of another. However, our methods could equally-well be used to investigate
other variations of parameters.
We believe that the numerical and analytical methods
of Sec.~\ref{sec:phen} and Sec.~\ref{sec:bif} can be extended
to other types of networks consisting of QIF neurons,
for which stationary, oscillating and moving patterns
also define their typical dynamical behaviour.
In particular, these methods can be useful for studying two-, three-
and higher-dimensional arrays, networks with population structure
and networks with propagation delays.
Even for the model considered above, it is possible to further investigate which patterns will occur when using other coupling functions $W_\mathrm{s}(x)$ and $W_\mathrm{v}(x)$, e.g.~wrapped Gaussian, exponentially decaying functions, etc.
Moreover, it would be interesting to check if these patterns persist in the presence of external noise. In particular, our methods may be able to be generalized to take advantages of the ``pseudocumulant'' expansions proposed in~\cite{goldiv21}.
We leave these questions for future work.

\section*{Appendix}

Let us consider a complex Riccati equation
\begin{equation}
\df{w}{t} = \gamma + i f(t) - g(t) w - i w^2
\label{Eq:Riccati:w}
\end{equation}
with a parameter $\gamma\in\mathbb{R}$ and real-valued functions $f(t)$ and $g(t)$.
In this section, we show that for $\gamma > 0$
all solutions of Eq.~(\ref{Eq:Riccati:w})
have special invariance properties with respect to the complex half-plane
$$
\mathbb{P} = \{ w\in\mathbb{C}\::\: \Real\:w> 0 \}. 
$$

In order to proceed, we recall that the complex plane transformation
\begin{equation}
z = \fr{1 - w}{1 + w}
\label{w_to_z}
\end{equation}
maps the half-plane~$\mathbb{P}$ onto the unit disc
$$
\mathbb{D} = \{ z\in\mathbb{C}\::\: |z| < 1 \}. 
$$
Moreover, using formula~(\ref{w_to_z}) as a change of variable,
we can rewrite Eq.~(\ref{Eq:Riccati:w}) in an equivalent form
\begin{equation}
\df{z}{t} = - \fr{ \gamma + i f(t) }{2} ( 1 + z )^2 + \fr{g(t)}{2} (1 - z^2) + \fr{i}{2} (1 - z)^2.
\label{Eq:Riccati:w_}
\end{equation}
Then, it is easy to see that Eq.~(\ref{Eq:Riccati:w_}) is a complex Riccati equation
$$
\df{z}{t} = c_0(t) + c_1(t) z + c_2(t) z^2
$$
with
$$
c_0(t) = \fr{ -\gamma - i f(t) + g(t) + i }{2},\quad
c_1(t) = -\gamma - i f(t) - i,\quad
c_2(t) = \fr{ -\gamma - i f(t) - g(t) + i }{2}.
$$
Therefore, Proposition~2 from~\cite{LaiO2023} implies
that for every $\gamma > 0$ and for all real $2\pi$-periodic continuous functions $f(t)$ and $g(t)$,
Eq.~(\ref{Eq:Riccati:w_}) has exactly one stable $2\pi$-periodic solution
and this solution satisfies $|z(t)| < 1$ for all $0\le t \le 2\pi$.
Moreover, using the main argument of the proof of Proposition~2 in~\cite{LaiO2023},
we can show that every solution of Eq.~(\ref{Eq:Riccati:w_})
with an initial condition in~$\mathbb{D}$ remains in this unit disc for any finite time interval.

Due to the equivalence between the solutions of Eq.~(\ref{Eq:Riccati:w})
and the solutions of Eq.~(\ref{Eq:Riccati:w_}), we can reformulate the above results as follows.

\begin{proposition}
Suppose that $\gamma > 0$ and $f(t)$, $g(t)$ are real $2\pi$-periodic continuous functions.
\smallskip

(i) Then for every $w_0\in\mathbb{P}$
the complex Riccati equation~(\ref{Eq:Riccati:w}) with initial condition $w(0) = w_0$
has a solution $w(t)$, which is bounded on the interval $[0,2\pi]$ and satisfies $\Real\:w(t) > 0$.
\smallskip

(ii) Moreover, Eq.~(\ref{Eq:Riccati:w}) has exactly one stable $2\pi$-periodic solution
and this solution satisfies $\Real\:w(t) > 0$ for all $0\le t \le 2\pi$.
\label{Proposition:Sln:w}
\end{proposition}

Since Eq.~(\ref{Eq:Riccati:w}) belongs to the class of complex Riccati equations
its Poincar{\'e} map coincides with a M{\"o}bius transformation, see~\cite{Cam1997,Wil2008}.
This fact can be used to speed up the computation of periodic solutions of Eq.~(\ref{Eq:Riccati:w}).

\begin{remark}
\label{remark2}
If the conditions of Proposition~\ref{Proposition:Sln:w} are fulfilled,
then the stable solution of Eq.~(\ref{Eq:Riccati:w}) can be computed in the following way.
\smallskip

(i) One solves Eq.~(\ref{Eq:Riccati:w}) on the interval $t\in(0,2\pi]$
with three different initial conditions $w(0) = w_k\in\mathbb{P}$, $k=1,2,3$,
and obtains three solutions $W_k(t)$.
Since each $w_k$ lies in the half-plane~$\mathbb{P}$
this automatically implies that $W_k(t)$ is bounded on the interval $t\in[0,2\pi]$.
Note that this fact is not obvious a priori,
since some of the solutions of Eq.~(\ref{Eq:Riccati:w}) blow up in finite time.
\smallskip

(ii) One denotes $\zeta_k = W_k(2\pi)$. Then, due to the properties
of Poincar{\'e} map one has $\zeta_k = \mathcal{M}(w_k)$, $k=1,2,3$,
where $\mathcal{M}(w)$ is a M{\"o}bius transformation representing this map.
The above three relations can be used to reconstruct the map $\mathcal{M}(w)$, namely
$$
\mathcal{M}(w) = \frac{a w + b}{c w + d}
$$
where
\begin{eqnarray*}
&&
a = \det \left(
\begin{array}{ccc}
 w_1 \zeta_1 & \zeta_1 & 1 \\[2mm]
 w_2 \zeta_2 & \zeta_2 & 1 \\[2mm]
 w_3 \zeta_3 & \zeta_3 & 1
\end{array}
\right),
\quad
b = \det \left(
\begin{array}{ccc}
 w_1 \zeta_1 & w_1 & \zeta_1 \\[2mm]
 w_2 \zeta_2 & w_2 & \zeta_2 \\[2mm]
 w_3 \zeta_3 & w_3 & \zeta_3
\end{array}
\right),
\\[2mm]
&&
c = \det \left(
\begin{array}{ccc}
 w_1 & \zeta_1 & 1 \\[2mm]
 w_2 & \zeta_2 & 1 \\[2mm]
 w_3 & \zeta_3 & 1
\end{array}
\right),
\quad\phantom{w_1}
d = \det \left(
\begin{array}{ccc}
 w_1 \zeta_1 & w_1 & 1 \\[2mm]
 w_2 \zeta_2 & w_2 & 1 \\[2mm]
 w_3 \zeta_3 & w_3 & 1
\end{array}
\right).
\end{eqnarray*}
\smallskip

(iii) Once the map $\mathcal{M}(w)$ is known, one can find its fixed points by
solving the quadratic equation
$$
c w^2 + d w - a w - b = 0.
$$
This yields two roots
$$
w_\pm = \fr{a - d \pm \sqrt{ (a - d)^2 + 4 b c }}{2 c}.
$$
\smallskip

(iv) Choosing from the roots $w_+$ and $w_-$ the one
that lies in the half-plane $\mathbb{P}$,
one obtains the initial condition that determines the periodic solution of interest.
The latter can be computed by solving Eq.~(\ref{Eq:Riccati:w}) with this initial condition.
\smallskip

(v) Sometime it may happen that the Poincar{\'e} map $\mathcal{M}(w)$
is strongly contracting so that
$$
|\zeta_1 - \zeta_2 | + |\zeta_3 - \zeta_2 | < 10^{-8},
$$
where the value $10^{-8}$ is chosen through experience.
In this case, the calculations in steps~(ii) and~(iii) become inaccurate.
Then the initial condition of the periodic solution of interest
is approximately given by the average $(\zeta_1 + \zeta_2 + \zeta_3) /3$.
\smallskip

The above steps (i)--(v) can be understood as a constructive definition
of the solution operator of Eq.~(\ref{Eq:Riccati:w}),
which for every $\gamma > 0$ and all real $2\pi$-periodic coefficients $f(t)$ and $g(t)$
yields the corresponding stable $2\pi$-periodic solution of Eq.~(\ref{Eq:Riccati:w}).
%More detailed justification of this definition can be found in~\cite[Section~4]{ome23}.
%The algorithm in Sec.~\ref{Sec:Moebius} consists of applying the procedure above at every
%point on the spatial grid (in parallel).
\label{Remark:Sln:Operator}
\end{remark}

\begin{proposition}
For constant coefficients $f$ and $g$,
Eq.~(\ref{Eq:Riccati:w}) has a single fixed point satisfying $\Real(w) \ge 0$.
This fixed point is determined by the formula
\begin{equation}
w = \mathcal{S}( f, g, \gamma ) := \fr{1}{2} \sqrt{ \fr{\sqrt{ ( g^2 - 4 f )^2 + 16 \gamma^2 }
- g^2 + 4 f}{2} } + \fr{i}{2} \left( g -
\sqrt{ \fr{\sqrt{ ( g^2 - 4 f )^2 + 16 \gamma^2 }
+ g^2 - 4 f}{2} } \right).
\label{Eq:FP:w}
\end{equation}
\label{Proposition:FP:General}
\end{proposition}

{\bf Proof:}
Any fixed point of Eq.~(\ref{Eq:Riccati:w}) satisfies
$$
\gamma + i f - g w - i w^2 = 0.
$$
This quadratic equation has two solutions
\begin{equation}
w = - \fr{1}{2 i}\left( g \pm \sqrt{ g^2 + 4 i ( \gamma + i f ) } \right) = \fr{i}{2}\left( g \pm \sqrt{ g^2 - 4 f + 4 i \gamma } \right).
\label{Eq:UniformSln}
\end{equation}
Since $\gamma > 0$, de Moivre's formula yields
$$
g^2 - 4 f + 4 i \gamma = \sqrt{ ( g^2 - 4 f )^2 + 16 \gamma^2 } \left( \cos\left( \mathrm{arccot}\fr{g^2 - 4 f}{4\gamma} \right)
+ i \sin\left( \mathrm{arccot}\fr{g^2 - 4 f}{4\gamma} \right) \right).
$$
Hence
$$
\sqrt{g^2 - 4 f + 4 i \gamma} = \sqrt[4]{ ( g^2 - 4 f )^2 + 16 \gamma^2 } \left( \cos\left( \fr{1}{2}\mathrm{arccot}\fr{g^2 - 4 f}{4\gamma} \right)
+ i \sin\left( \fr{1}{2} \mathrm{arccot}\fr{g^2 - 4 f}{4\gamma} \right) \right).
$$
Using the half-angle formulas and the identity
$$
\cos\left( \mathrm{arccot}\fr{g^2 - 4 f}{4\gamma} \right) = \fr{g^2 - 4 f}{\sqrt{ ( g^2 - 4 f )^2 + 16 \gamma^2 }},
$$
we rewrite this as follows
$$
\sqrt{g^2 - 4 f + 4 i \gamma} = 
\sqrt{ \fr{\sqrt{ ( g^2 - 4 f )^2 + 16 \gamma^2 }
+ g^2 - 4 f}{2} }
+ i \sqrt{ \fr{\sqrt{ ( g^2 - 4 f )^2 + 16 \gamma^2 }
- g^2 + 4 f}{2} }.
$$
Inserting this expression into~(\ref{Eq:UniformSln})
and choosing the sign of the square root
that ensures the inequality $\Real(w) \ge 0$,
we obtain formula~(\ref{Eq:FP:w}).~\qed

%\nolinenumbers

\section*{Acknowledgements}
The work of O.E.O. was supported by the Deutsche Forschungsgemeinschaft
under Grant No. OM 99/2-3.

%\bibliographystyle{unsrt}
%\bibliography{spiral}

\end{document}